\documentclass[10pt,twocolumn]{revtex4}
\usepackage{graphicx} 
\usepackage{dcolumn}
\usepackage{bm}
\usepackage{hyperref}
\pdfoutput=1
\usepackage{amssymb} 
\usepackage{amsmath}
\usepackage{color}

\newcommand{\BE}{\begin{equation}}
\newcommand{\EE}{\end{equation}}
\newcommand{\BA}{\begin{eqnarray}}
\newcommand{\EA}{\end{eqnarray}}
\newcommand{\bx}{{\bf x}}
\newcommand{\by}{{\bf y}}

\newcommand{\bk}{{\bf k}}
\newcommand{\bq}{{\bf q}}
\newcommand{\bH}{{\bf H}}
\newcommand{\bG}{\textrm{G}}
\newcommand{\bL}{\textrm{L}}
\newcommand{\hr}{{\hat \rho}}
\newcommand{\hv}{\hat{v}}
\newcommand{\bxi}{{\boldsymbol\xi}}
\newcommand{\bet}{{\boldsymbol\eta}}

\newcommand{\tD}{\tilde{D}}
\newcommand{\tv}{\tilde{v}}

\begin{document}

\title{
Pattern formation with repulsive soft-core interactions:
discrete particle dynamics and Dean-Kawasaki equation}

\author{Jean-Baptiste Delfau}
\affiliation{IFISC (CSIC-UIB), Instituto de F\'{\i}sica
Interdisciplinar y Sistemas Complejos, E-07122 Palma de
Mallorca, Spain}

\author{H\'{e}l\`{e}ne Ollivier}
\affiliation{\'Ecole Normale Sup\'{e}rieure de Cachan, 94230 Cachan, France }
\affiliation{IFISC (CSIC-UIB), Instituto
de F\'{\i}sica Interdisciplinar y Sistemas Complejos, E-07122
Palma de Mallorca, Spain}

\author{Crist\'{o}bal L\'opez}
\affiliation{IFISC (CSIC-UIB), Instituto de F\'{\i}sica
Interdisciplinar y Sistemas Complejos, E-07122 Palma de
Mallorca, Spain}

\author{Bernd Blasius}
\affiliation{Institute for Chemistry and Biology of the Marine
Environment, University of Oldenburg,
Carl-von-Ossietzky-Strasse 9-11, 26111 Oldenburg, Germany}

\author{Emilio Hern\'andez-Garc\'{\i}a}
\affiliation{IFISC (CSIC-UIB), Instituto de F\'{\i}sica
Interdisciplinar y Sistemas Complejos, E-07122 Palma de
Mallorca, Spain}

\date{\today}

%
\begin{abstract}
Brownian particles interacting via repulsive soft-core
potentials can spontaneously aggregate, despite repelling each
other, and form periodic crystals of particle clusters. We
study this phenomenon in low-dimensional situations (one and
two dimensions) at two levels of description: performing
numerical simulations of the discrete particle dynamics, and by
linear and nonlinear analysis of the corresponding
Dean-Kawasaki equation for the macroscopic particle density.
Restricting to low dimensions and neglecting fluctuation
effects we gain analytical insight into the mechanisms of the
instability leading to clustering which turn out to be the
interplay between diffusion, the intracluster forces and the
forces between neighboring clusters. We show that the
deterministic part of the Dean-Kawasaki equation provides a
good description of the particle dynamics, including width and
shape of the clusters, in a wide range of parameters, and
analyze with weakly nonlinear techniques the nature of the
pattern-forming bifurcation in one and two dimensions. Finally,
we briefly discuss the case of attractive forces.
\end{abstract}

\maketitle


\section{Introduction}
\label{sec:introduction}

Ensembles of interacting random walkers and
their description in terms of densities appear in many contexts
ranging from biological or physical to social phenomena
\cite{Grimm2005,Bonabeau2002,Topaz2006,Chavanis2008a,Marchetti2013,Pineda2010}.
Usually these interactions act locally, involving only a few
individuals, but they induce global patterns of behavior of the
full system like phase transitions, the formation of periodic
spatial structures, collective movement and synchronization
states.
Knowing the conditions for the formation of these collective
structures and its own feedback on the dynamics is a central
issue in the understanding of complex systems. The study of
these individual-based models is approached from two
complementary points of views: a) the particle description,
describing the dynamics of individuals and their interactions,
and based mainly on numerical simulations; and b) the continuum
description in terms of evolution equations for the local
density (or another macroscopic field).

In physical systems forces drive particle motions, and they are
usually derived from two-body potentials acting repulsively or
attractively at different distances. This is the case of many
forces, like the Lennard-Jones case, among atoms and molecules
in liquids, polymer and colloidal solutions. In biological
systems facilitation and competition mechanisms at short and
large scales also drive organism motion, but they can in
addition modulate growth and death rates \cite{Murray2004,Okubo2001}. 
The interplay between
facilitation and competition at different distances, but
specially the effect of competition, has been shown to be
responsible for the formation of periodic arrangements of
clusters of particles and more complex structures
\cite{Hernandez2004,Hernandez2014,Schellinck2011,MartinezGarcia2015},
which are related to the appearance of vegetation patterns and
periodic aggregations of bacteria
\cite{Borgogno2009,Martinez2014,BenJacob2000,Ramos2008,Rietkerk2008,Meron}.

There is however a recently discovered situation, relevant to
polymer and colloidal solutions, where the same effect is
observed for particle systems interacting with soft-core forces
which are repulsive at all distances
\cite{Klein1994,Likos2001,Mladek2006,Likos2007,Mladek2008,Coslovich2013}:
a liquid-solid transition occurs, but in the solid the unit
cell is not occupied by one particle or molecule, but by a
closely packed cluster of them, forming a so-called cluster
crystal. Note the counter-intuitiveness of this phenomenon:
{\it despite all particles are repelling, they aggregate}.
Beyond condensed-matter systems this is a phenomenon analogous
to the aggregation of reproducing organisms occurring despite
purely competitive interactions
\cite{Hernandez2004,MartinezGarcia2013,Hernandez2014}.

This cluster crystallization transition has been analyzed with
equilibrium statistical mechanics tools, including Monte-Carlo
simulations, in three-dimensional systems
\cite{Likos2001,Mladek2006,Likos2007,Mladek2008,Coslovich2013}.
However, approaches exploiting non-linear
dynamics and pattern formation techniques \cite{Walgraef2012}
can add insight to the study of this instability, including
dynamic regimes. This was in fact the route followed in the
early work by Munakata \cite{Munakata1977}, but the lack of
numerical simulations there hindered the identification of
several relevant features, as for example the dominance of
hexagonal patterns instead of stripes in two dimensions. In
addition, identification and understanding of relevant
mechanisms become much clearer when considering low-dimensional
(one- and two-dimensional) systems, as compared with the
complexity of three-dimensional structures. The central
objective of this paper is to analyze in detail the processes
leading to cluster crystals in one-dimensional (1d) and
two-dimensional (2d) systems of soft-core repulsive particles
providing when possible analytical insight. The physical
mechanisms leading to this cluster formation are discussed in
detail. Our approach is similar in spirit to the one used in
\cite{Archer2013,Archer2015}, where 2d crystals arising from
competing interactions at two spatial scales were considered,
but here we restrict to the simpler case of a single
interaction scale for which greater and more complete
understanding can be achieved. For completeness, we consider
briefly also the case of purely attracting interactions,
highlighting some similarities and differences between both
cases.

The paper is organized as follows: In Section
\ref{sec:particlerepulsive} we start our study with an
overdamped Brownian dynamics, which is relevant in freezing,
the glass transition, colloidal systems, or bacterial patterns,
and investigate the effect of repulsive interactions, leading
to cluster crystals, by performing numerical simulations of the
particle dynamics.  In Sec. \ref{sec:DKrepulsive} we analyze a
deterministic integrodifferential model, the Dean-Kawasaki (DK)
equation \cite{Kawasaki1994,Dean1996,Munakata1977,Kim2014}, showing
that it gives an appropriate description of the particle
dynamics and give analytical arguments for the findings of the
previous section. In particular we obtain analytical results
for the pattern formation transition and for the shape of the
pattern and of the clusters forming it. For completeness, we
briefly consider attractive interactions in Sect.
\ref{sec:attractive} both for the particles and for the DK
model. Finally, in Sec. \ref{sec:summary} we give a discussion
and summary of the main results.


\section{Brownian particles with soft-core repulsive interactions}
\label{sec:particlerepulsive}

\begin{figure*}
\includegraphics[width=\textwidth]{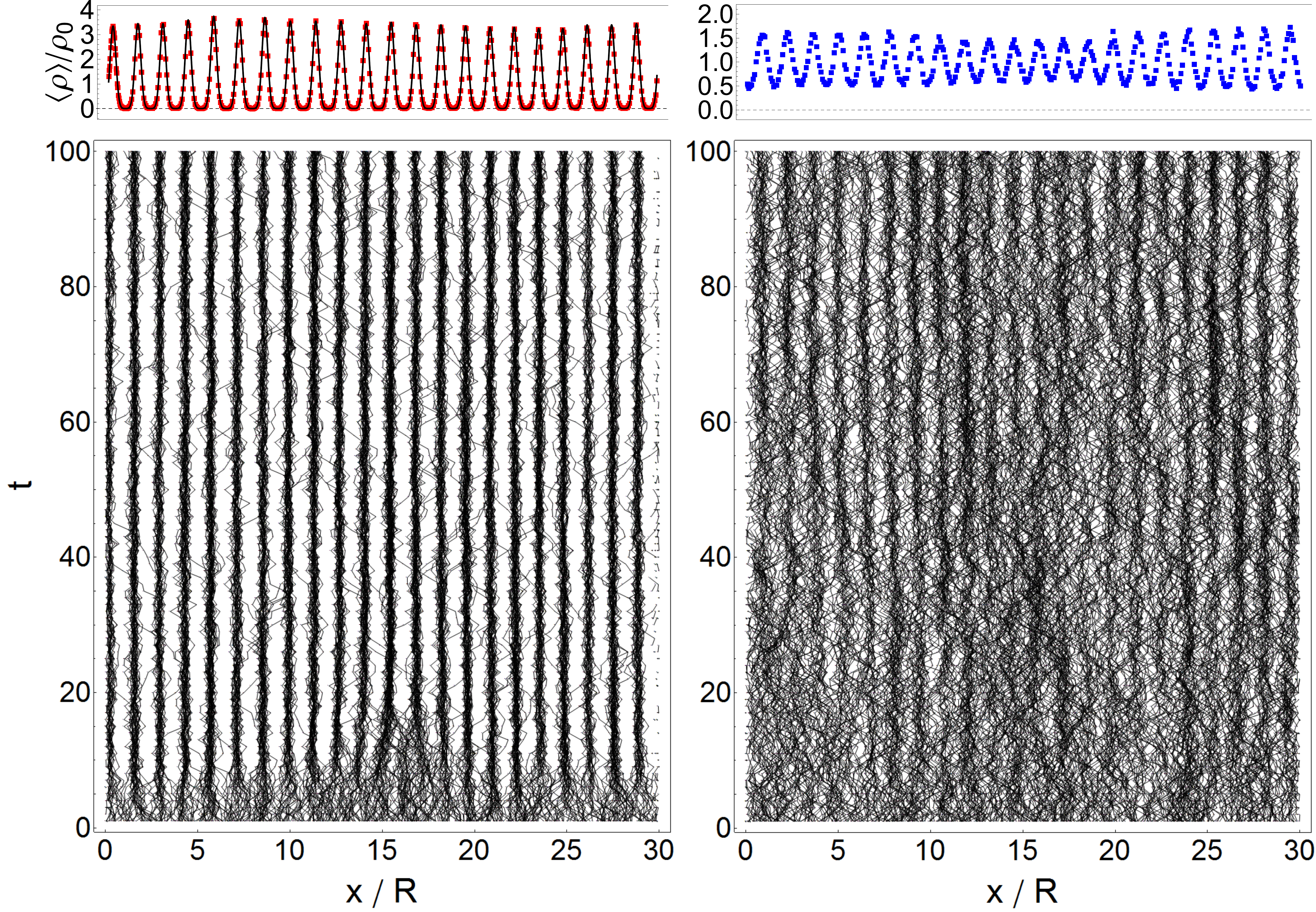}
\caption{\label{fig:1Dparticles} Dynamics of a system of 6000
particles evolving according to Eq.~(\ref{Brownian}) in a
periodic one-dimensional domain of size $L=3$, so that $\rho_0
= 2000$. The interaction potential is GEM-3 with parameters $R
= 0.1$ and $\epsilon=0.0333$. Left, $D=0.4$ so that $\tD =
D/(\epsilon\rho_0 R)=0.06$. Right:  $D=0.6$ so that $\tD=0.09$.
The lower panels show the spatiotemporal trajectories (in
black). For clarity of the plots, only $600$ trajectories are
shown here. The top panels show a coarse-grained normalized
density, $\left<\rho(x)\right> / \rho_0$, at the late stages.
The coarse-graining is done by averaging spatially over boxes
of width $0.02 R$ and temporally over the last $150$ temporal
configurations separated by $10^{-3}$ time units. In the top
left panel, the clusters are fitted by Gaussians (black
lines).}
\end{figure*}

In the overdamped limit, the motion of $N$ point particles at
positions $\{\bx_1,\bx_2,...,\bx_n\}$ in $d$-dimensional space,
with friction coefficient $\gamma$, subjected to Brownian
motion and interacting with a potential energy $U=\gamma V$ is
given by
\BE
\dot\bx_i = - \nabla_i V(\bx_1,\bx_2,...,\bx_n)
+\sqrt{2D}~\bxi_i(t) \ , \label{Brownian0}
\EE
with independent Gaussian noise vectors $\bxi_i$ satisfying
\BE
\left<\bxi_i\right>=0 \ , \
\left<\bxi_i(t)\bxi_j(t')\right>=\mathbb{I}\delta_{ij}\delta(t-t')
\ .
\EE
$\mathbb{I}$ is the $d$-dimensional identity. When noise is of
thermal origin the diffusion coefficient $D$ is proportional to
temperature according to Einstein's relation $D=k_B T /\gamma$.
$\nabla_i$ denotes the gradient with respect to the position
$\bx_i$. The system's mean density is given by $\rho_0 = N /
L^d$  ($L^d$ the total $d-$volume) and remains constant since
the total number of particles is conserved. In the following we
will assume pairwise interactions, so that $V(\bx_1,...\bx_N)=
\frac{1}{2}\sum_{ij} v(\bx_i- \bx_j)$, and (\ref{Brownian0})
becomes
\BE
\dot\bx_i = - \sum_{j=1}^N\nabla v(\bx_i-\bx_j)
+\sqrt{2D}~\bxi_i(t).
\label{Brownian}
\EE

\begin{figure}
\includegraphics[width=.96\columnwidth]{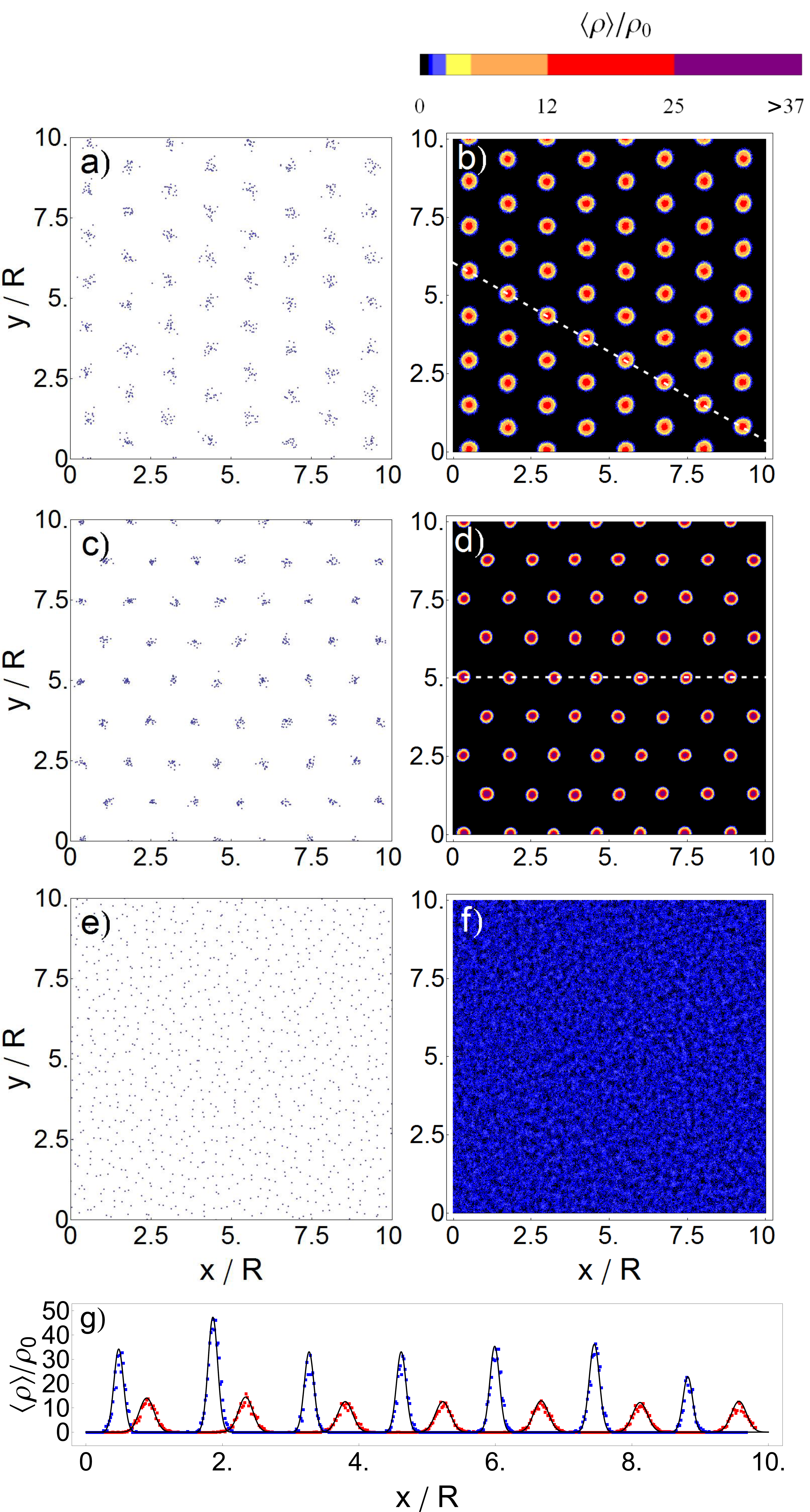}
\caption{\label{fig:P_simu}Left column: snapshots of the positions of $N=1000$ particles in
2d at large times for $R=0.1$, $\epsilon = 0.0333$, $L=1$ and $\rho_0 = 1000$.
Right column: coarse-grained densities of the same
configurations. The coarse-graining is done by averaging in space over distances $0.017 R$ and
in time over $500$ configurations separated by $10^{-3}$ time units. a) and b) GEM-3 potential with
$D = 0.02$ so that $\tD = D/(\epsilon\rho_0 R^2)=0.06$. c) and d) GEM-3 potential
with $D = 0.01$ ($\tD =0.03$). e) and f) GEM-1 potential with $D = 0.005$ ($\tD =0.015$). Figure g) is
the density $\rho$ along the white dotted lines shown on Figs. b) and d) (red and blue
squares respectively). The black curves correspond to a fit by a sum of Gaussian functions.}
\end{figure}

Following previous works \cite{Mladek2006,Likos2007} we
consider the generalized exponential model of exponent $\alpha$
(GEM-$\alpha$) as interaction potential:
\BE
v(\bx) \equiv \epsilon
\exp\left(-\left|\frac{\bx}{R}\right|^\alpha\right) \ .
\label{GEM_def}
\EE
This is a convenient and flexible family of interactions
sharing the property of {\sl soft-core} that will be relevant
for the cluster crystallization. By {\sl soft-core} we mean
that the potential does not diverge at $\bx=\bf 0$ so that the
particles can overlap. The width $R$ indicates the spatial
range of the interaction, and $\epsilon$ its magnitude.
$\epsilon$ is positive for the repulsive interactions mostly
considered here, and will be taken to be negative in Sect.
\ref{sec:attractive} to model attractive interactions. For
$\alpha <2$ GEM-$\alpha$ potentials are more peaked at zero,
and they get more box-like as $\alpha$ increases. It is known
\cite{Bochner1937,Likos2007} that GEM-$\alpha$ potentials with
$\alpha>2$ have both positive and negative Fourier components,
while for $\alpha \leq 2$ the Fourier transform is strictly
positive. These will be important properties as we will see
later. In our simulations two representative kinds of GEM
potentials are used: a GEM-1 and a GEM-3 potential, although
results for other values of $\alpha$ will be mentioned.

Numerically we observe that for large diffusion coefficients
or any  GEM-$\alpha$ potential with $\alpha <2$ the particle
distribution remains rather unstructured in time. For
GEM-$\alpha$ with $\alpha>2$, however, a periodic modulation in
the particle distribution arises when decreasing the diffusion
coefficient or increasing the density. This is seen in Figure \ref{fig:1Dparticles}, that
shows spatiotemporal trajectories of a one-dimensional system of
Brownian particles moving according to Eq.~(\ref{Brownian})
with repulsive interaction GEM-3, starting from an initial
random distribution. For sufficiently small $D$, particle
distribution becomes a periodic array of well separated clusters.
This is a cluster crystal as each cluster is made of many
particles which remain very close, despite they repel each
other. Also in Fig. \ref{fig:1Dparticles} we show a
coarse-graining of the particle distribution at the latest
times, showing that the configuration is essentially an array
of Gaussian clusters for small $D$, approaching a sinusoidal
modulation when increasing $D$ towards the disappearance of the
pattern. By slowly incrementing $D$ in our particles simulations, 
we determined that this occurs at $D_c \approx 0.68$ for the parameters
used in that figure.

In Figure \ref{fig:P_simu}, we plot results from 2d simulations of
the particle system under the GEM-3 and the GEM-1 potentials.
Left column shows long-time distributions of the particles for
these two interacting potentials and for different values of
the control parameters, while in the right column we show
the coarse-grained density functions. It clearly appears on
Fig.  a) and c) that hexagonal patterns can spontaneously
appear for the GEM-3 potential. The corresponding densities in
the right column exhibit peaks that can be fitted by
two-dimensional Gaussians as can be seen in Fig.
~\ref{fig:P_simu} g). The width of the peak $\sigma$ and the
distance $a$ between them are functions of the control
parameters as we will see later on. For a given GEM-3
potential, these patterns disappear when $D$ is increased, or
when either $\rho_0$, $R$ or $\epsilon$ are decreased. On the
other hand, persistent clusters are never observed for
particles interacting with the GEM-1 potential. In those cases,
the late time state is always statistically homogeneous as in
Fig. ~\ref{fig:P_simu} e), f).

\begin{figure*}
\includegraphics[width=.32\textwidth]{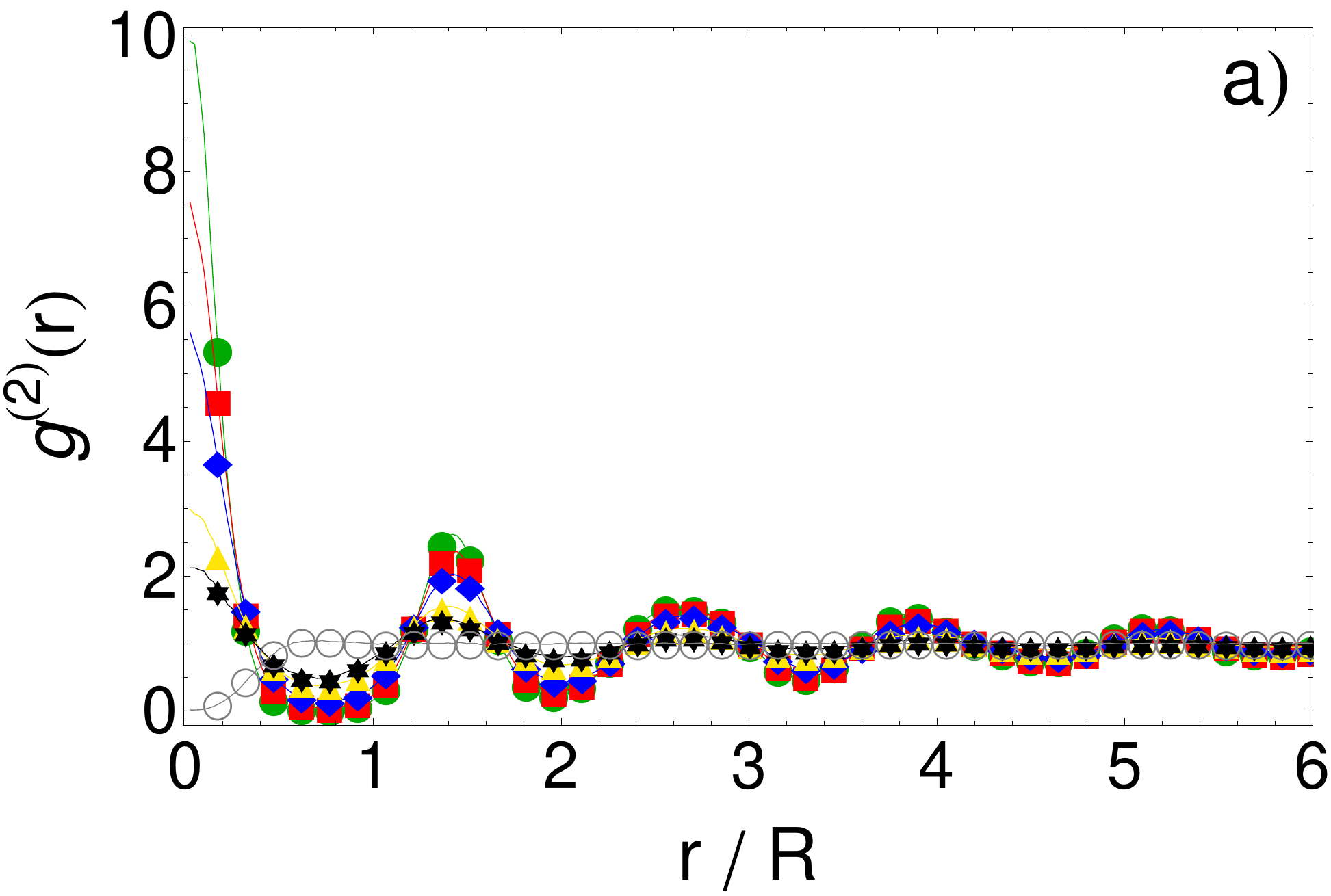}
\includegraphics[width=.32\textwidth]{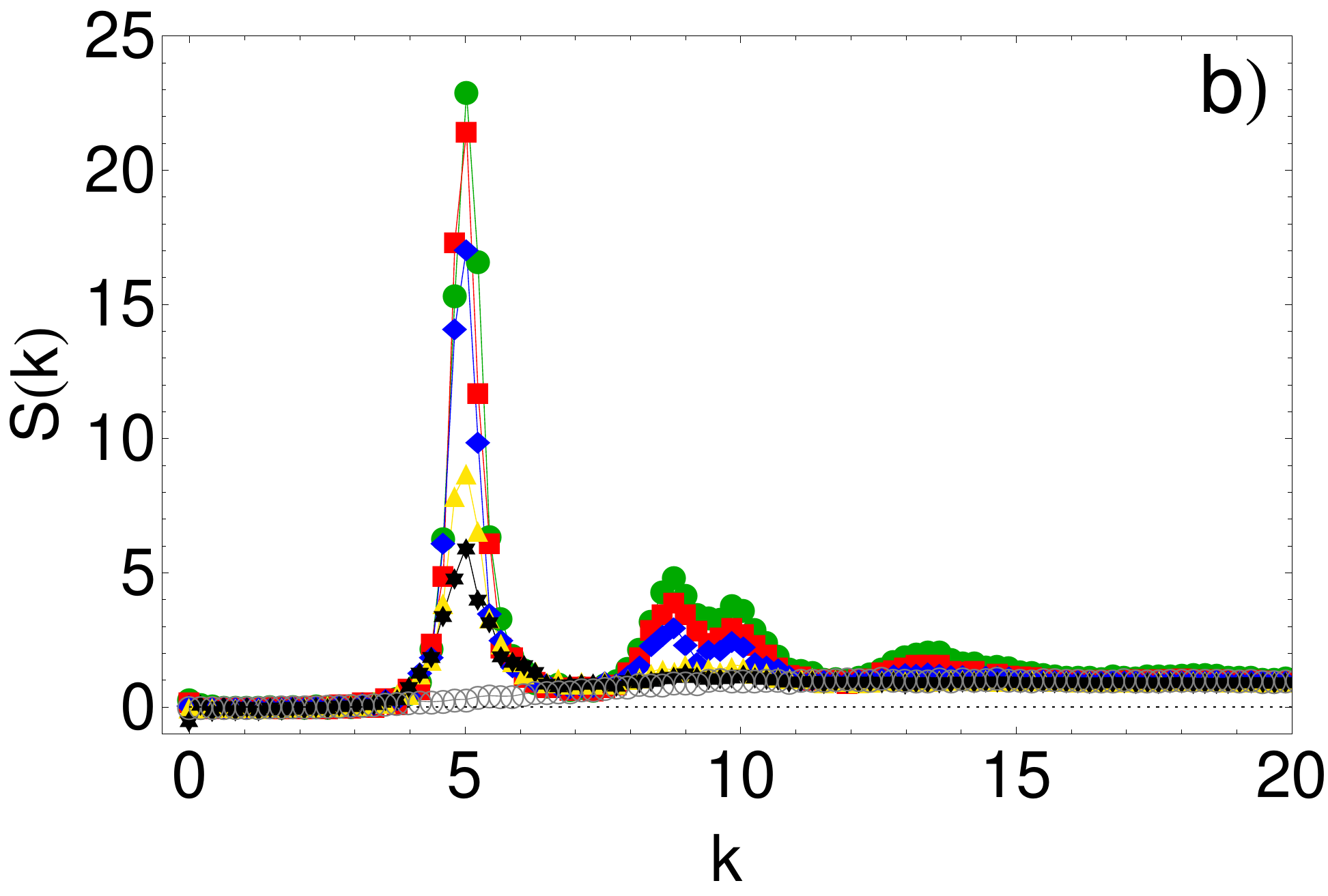}
\includegraphics[width=.32\textwidth]{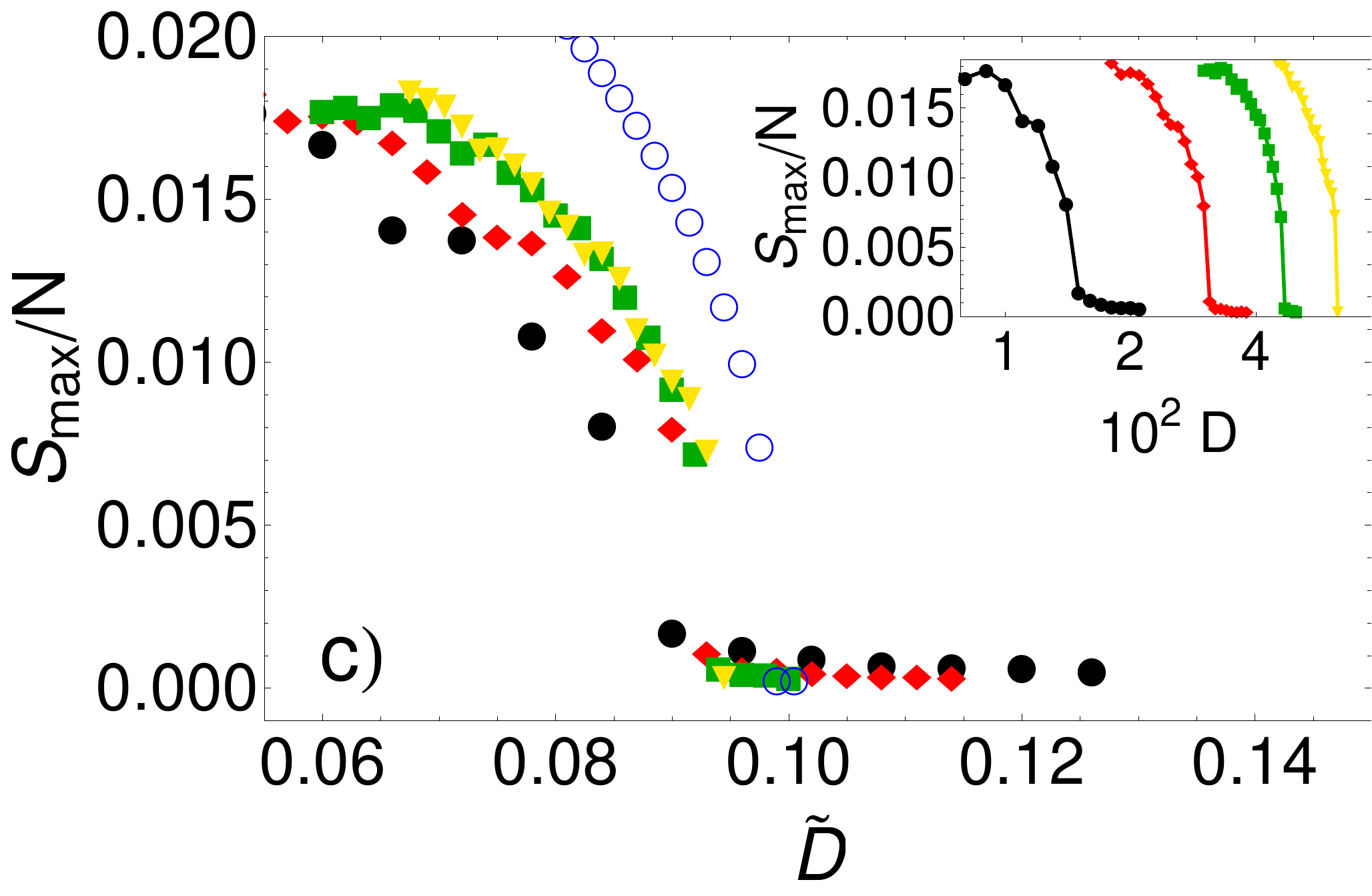}
\caption{\label{fig:structure}a) Radial distribution function
$g^{(2)}(r)$ and b) structure factor $S(k)$ (as a function of
the dimensionless wavevector $k=qR$, where $q$ is the Fourier variable 
in the definition of $S(q)$ in Eq.~(\ref{Sq})) for several values of
$\tilde{D}=D/(\epsilon\rho_0R^2)$ in 2d particle simulations
for $L = 3$, $\epsilon = 0.0333$ and $R = 0.1$. GEM-3 potential
and $\tilde{D} = 0.05$ (green disks), $0.06$ (red squares),
$0.07$ (blue diamonds), $0.08$ (yellow triangles), $0.09$
(black stars) and GEM-1 potential with $\tilde{D} = 0.05$ (gray
circles). c) Normalized maximum value, $S_{max}/N$, of the main peak of
the structure factor plotted with respect to $D$ (inset) or $\tilde{D}$ (main
plot) in 2d particle simulations for $L = 9$, $\epsilon =
0.0333$ and $R = 0.1$. Yellow triangles : $\rho_0 = 2000$.
Green squares : $\rho_0 = 1500$. Red diamonds : $\rho_0 =
1000$. Black disks : $\rho_0 = 500$. The blue circles
correspond to results obtained by integration of the 2d
deterministic DK equation for the same parameters and $\rho_0 =
2000$. In that case, $S(k)$ was obtained by taking the Fourier transform of the density-density correlation function.}
\end{figure*}

The structure of the system is conveniently analysed
by computing the radial distribution function $g^{(2)}(r)$ and
the structure factor $S(q)$ given  by
\BA
g^{(2)}(\bx) & = & g^{(2)}(r) = \frac{1}{\rho_0} \langle
\sum_{i \neq 0} \delta(\bx-\bx_i)\rangle,
\label{gr}\\
S(\bq) & = & S(q) =  1 + \rho_0 \int d \bx e^{-i \bq \cdot
\bx}\left(g^{(2)}(\bx) - 1\right) \hspace{-0.1cm}.
\label{Sq}
\EA
The sum in (\ref{gr}) is over particles different from a
reference particle at the origin, and the average is over positions
$\bx$ at the same distance $r=|\bx|$ from there.
Figure~\ref{fig:structure} a) shows that the radial
distributions of systems with GEM-3 potentials have several
peaks. The first one, at $r=0$, corresponds to particles
belonging to the same cluster while the second peak tells us
the typical distance $a$ between two neighboring clusters. The
height of the peaks is proportional to how ``ordered'' the
system is so that it decreases for larger diffusion
coefficients. It is also interesting to note that the $r=0$
peak completely disappears for the GEM-1 potential and can thus
be considered as a signature of the $\alpha>2$ clustering, in
which the ultra-soft potentials allow particles to concentrate
at very short distances. The structure factor $S(q)$, which can
be calculated from the Fourier transform of $g^{(2)}(r)$, also
has a clearly visible peak at the
wavelength $q=2\pi/a$. Its amplitude $S_{max}$ decreases when
$D$ increases as can be seen in Fig. ~\ref{fig:structure} c) with scaled units.
It also decreases for decreasing $R$, $\rho_0$ and $\epsilon$.
The change of $S_{max}$ with respect to $D$ underlines the
transition from periodic patterns to homogeneous equilibrium
states as it abruptly jumps down to $S_{max} \approx 1$ for a
critical value of $D_c$.

It should be noted that, according to Peierls argument
\cite{Peierls1935quelques}, we do not expect a true crystal
with long-range order in one-dimension if $D$ (proportional to
temperature) is non-zero. Also, although standard theorems on
absence of phase transitions do not apply to the soft-core
potential used here \cite{Cuesta2004}, it seems that true
thermodynamic phase transitions do not occur in this type of
models in one dimension \cite{Prestipino2015}. Thus, when
referring to states such as the ones displayed in Fig.
\ref{fig:1Dparticles} as \textsl{cluster crystals} we do not
imply the existence of any true thermodynamic solid-liquid
phase transition, but simply highlight that the local
organization of the particle distribution at small $D$ is very
different and more clustered and periodic than the nearly
homogeneous state found at large $D$. When neglecting
fluctuations, however, the transition becomes a true
bifurcation, as will be seen in Sect. \ref{sec:DKrepulsive}. It
will also be shown there that this deterministic approximation
gives a reasonable description if not too close to the
bifurcation point. The situation in 2d is more subtle, because
of the peculiarities of two-dimensional melting
\cite{Strandburg1988}. We will not address here the nature of
the crystal-liquid transition in this soft-core system
\cite{Gasser2009,Zu2016}. We just note that the approximate scaling of $S_{max}$ with $N=\rho_0 L^2$ (Fig. \ref{fig:structure}c) suggests the presence of some  translational order in the system. As in the 1d case, the deterministic approach described in the next section provides useful insight of the mechanisms at work and even quantitative description of the observations in some parameter range.

\section{Description in terms of the Dean-Kawasaki equation}
\label{sec:DKrepulsive}

Analytical arguments for the numerical results found in the
previous section can be derived from the continuum density
equation of the system of particles. This is given by the
Dean-Kawasaki (DK) equation \cite{Dean1996,Kawasaki1994}, which
is the following stochastic partial differential equation:
\BA
\partial_t \rho(\bx,t) = \nabla \cdot \left(\rho(\bx,t)\int d\bx' \nabla
v(\bx-\bx')\rho(\bx',t)\right)  \nonumber \\
+ D\nabla^2 \rho(\bx,t) + \nabla \cdot
\left(\sqrt{2D\rho(\bx,t)} ~\bet(\bx,t)\right) \ ,
\label{Dean}
\EA
with the spatio-temporal Gaussian noise vector satisfying
\BE
\left<\bet(\bx,t)\right>=0 \ , \
\left<\bet(\bx,t)\bet(\bx',t')\right>=\mathbb{I}\delta(\bx-\bx')\delta(t-t')
\ . \label{DeanNoise}
\EE
The noise and the diffusion terms arise from the random motion
of the Brownian particles, whereas the term containing the
potential describes the density advection by the local velocity
produced by the repulsion forces.

In the original Dean's derivation \cite{Dean1996},
$\rho(\bx,t)$ is the microscopic density
$\hr(\bx,t)=\sum_{i=1}^N \delta\left(\bx-\bx_i(t)\right)$, so
that the equation is of not much use, since it contains exactly
the same information as Eq.~(\ref{Brownian}) but in a much more
involved manner.
The deterministic version of the DK equation however is
affordable analytically and provides complementary
understanding. Dean's derivation \cite{Dean1996} uses \^{I}to
calculus, but the associated Stratonovich equation is indeed
the same because of the vanishing of the spurious drift
\cite{Gardiner} for the conserved noise in Eq.~(\ref{Dean}).
Alternatively, Kawasaki derivation~\cite{Kawasaki1994} shows
that the DK equation is also an approximation to the dynamics
of a coarse graining $\rho(\bx,t)$ of the microscopic density
$\hr(\bx,t)$, when the coarse graining of the product
$\hr(\bx,t)\hr(\bx',t)$ is approximated by
$\rho(\bx,t)\rho(\bx',t)$.

The gradient operator in all terms reflects the
particle-conserving character of the equation, so that the
total number of particles $N(t) \equiv \int d\bx \rho(\bx,t)$
does not change in time and remains always equal to the number
of particles $N$ of the particle description
Eq.~(\ref{Brownian}). If the initial density $\rho(\bx,t=0)$ is
non-negative everywhere, positivity is preserved in time. See
\cite{Archer2004} for further discussion on the meaning of the
DK equation and its relationship with dynamic density
functional theory~\cite{Marconi1999}.



We introduce dimensionless variables $\tilde{\bx} = \bx / R$,
$\tilde{t} = t \rho_0 \epsilon R^{d-2}$ and
$\tilde{\rho}(\tilde{\bx},\tilde{t})=\rho(\bx,t)/\rho_0$, so
that Eq.~(\ref{Dean}) becomes
\BA
\partial_{\tilde{t}} \tilde\rho(\tilde{\bx},\tilde{t}) =
\tilde\nabla \cdot
\left(\tilde{\rho}(\tilde{\bx},\tilde{t})\int d\tilde{\bx}'
\tilde\nabla
\tilde{v}(\tilde{\bx}-\tilde{\bx'})\tilde{\rho}(\tilde{\bx}',\tilde{t})\right)  \nonumber \\
+ \tilde{D}\tilde\nabla^2 \tilde{\rho}(\tilde{\bx},\tilde{t}) +
\frac{1}{\sqrt{n_R}}\tilde\nabla \cdot \left(\sqrt{2\tilde
D\tilde\rho(\tilde\bx,\tilde t)} ~\tilde\bet(\tilde\bx,\tilde
t)\right) \ , \label{Dean_nondim_stoch}
\EA
with the new spatio-temporal Gaussian noise vector satisfying
again
\BE
\left<\tilde\bet(\tilde\bx,\tilde t)\right>=0 \ , \
\left<\tilde\bet(\tilde\bx,\tilde
t)\tilde\bet(\tilde\bx',\tilde
t')\right>=\mathbb{I}\delta(\tilde\bx-\tilde\bx')\delta(\tilde
t-\tilde t') \ . \label{DeanNoise_nondim}
\EE
We have introduced the dimensionless potential
$\tv(\tilde\bx)=v(\bx)/\epsilon$, so that in our GEM-$\alpha$
case we have
\BE
\tv(\tilde\bx)=e^{-|\tilde\bx|^\alpha} \ .
\EE
Thus, besides the exponent $\alpha$ characterizing the
potential, we have just two relevant dimensionless parameters:
$\tilde{D} \equiv D / (\epsilon \rho_0 R^d)$, and $n_R \equiv
\rho_0 R^d$ (we assume system size $L/R$ to be sufficiently
large so that it would not play a relevant role). For
sufficiently large density or interaction range the parameter
$n_R$ will be also large and then the noise term would become
unimportant. The only remaining parameter will be then
$\tilde{D}$ which gives the ratio between the strength of
diffusion (or temperature) and of particle interactions. These
dimensional arguments explain the scaling with $\tD$ of the
different structure factors as observed in Fig.
~\ref{fig:structure} c): the values of $S_{max}$ approximately
collapse on the same curve when they are plotted with respect
to $\tD$. The small differences can be attributed to the
relative amplitude of the noise: when the density $\rho_0$
decreases $n_R$ decreases and the noise becomes stronger so
that the transition threshold shifts to smaller values of
$\tilde{D}$. In the rest of the paper we will neglect the noise
term and focus on the deterministic part of the DK equation.
We will see that this level of description provides good
results in some parameter range and, more importantly, allows
understanding of the mechanisms involved in the cluster
crystallization phenomenon.

\subsection{Pattern formation in the deterministic Dean-Kawasaki equation}
\label{subsec:patterrepulsiveDK}

\begin{figure}
\includegraphics[width=\columnwidth]{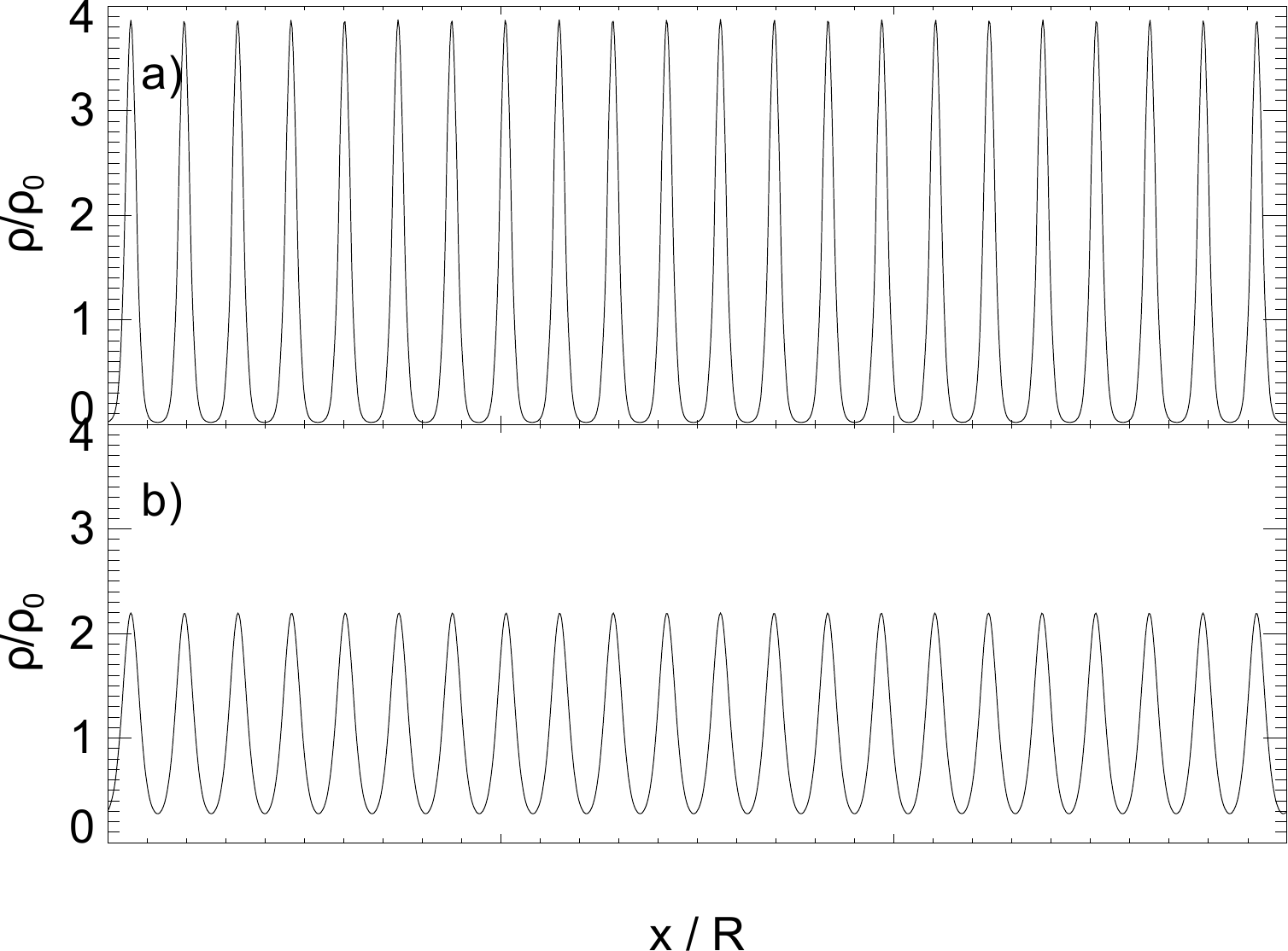}
\caption{\label{fig:Patterns_nonoise_1d} Steady solutions of Eq.~(\ref{Dean_nondim})
in 1d for $\alpha=3$. a) $\tD=0.06$. b) $\tD=0.09$ }
\end{figure}

To alleviate the notation in the following we will drop the
tilde on $\tilde{\bx}$, $\tilde{t}$ and $\tilde{\rho}$,
although it will be maintained on $\tD$ and $\tv$ to keep in
mind that they are dimensionless quantities. Eventually, some
results will be reverted back to the original variables, which
will then be referred as {\sl unscaled variables}. With this
notation, the deterministic part of the DK equation reads:
\BA
\partial_t \rho(\bx,t) &=& \nabla \cdot \left(\rho(\bx,t)\int d{\bx}' \nabla
\tilde{v}(\bx-\bx')\rho(\bx',t)\right)  \nonumber \\
&+& \tilde{D}\nabla^2 \rho(\bx,t)\ . \label{Dean_nondim}
\EA
Note that this dimensionless version is equivalent to using in
the original one, Eq.~(\ref{Dean}), the values
$R=\epsilon=\rho_0=1$. Equation (\ref{Dean_nondim}) has been
integrated numerically using standard pseudo-spectral methods.

Figure~\ref{fig:Patterns_nonoise_1d} shows a one-dimensional
configuration obtained at long times for the same parameters as
in Fig. ~\ref{fig:1Dparticles}. Besides the fact that the
coarse-grained particle densities are always more noisy than
the deterministic DK ones, the general agreement confirms that the deterministic description is
accurate enough. Also, density becomes homogeneous in the DK
simulations when increasing $\tD$, or for any $\tD$ if using a
GEM-$\alpha$ potential with $\alpha \le 2$. Note however that for
$\tD = 0.09$, the local density appears more sinusoidal for
particle simulations than for the DK equation.

\begin{figure}
\includegraphics[width=\columnwidth]{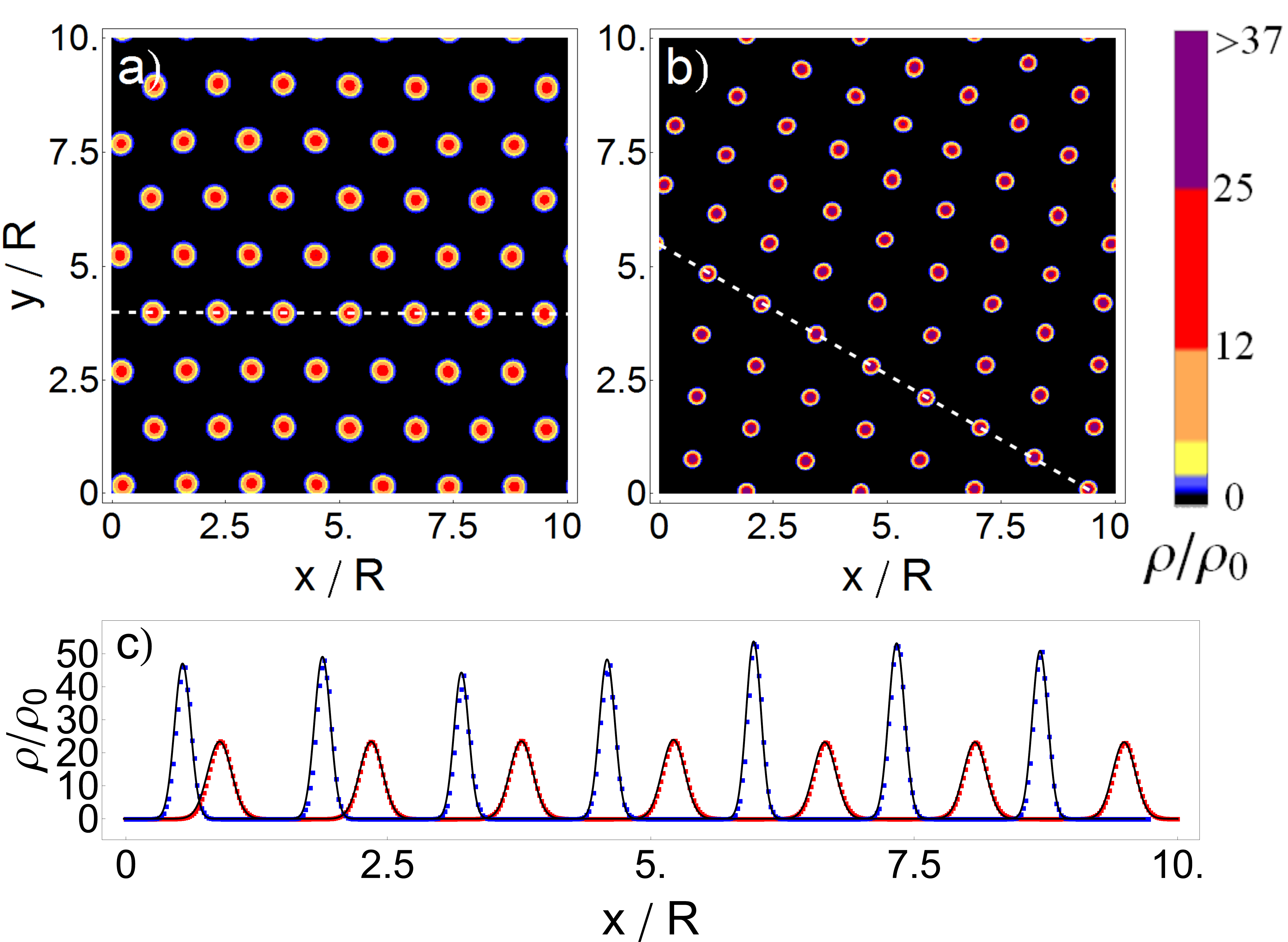}
\caption{\label{fig:Patterns_nonoise}Density functions in the steady state
for the GEM-3 potential obtained by integration of Eq.~(\ref{Dean_nondim})
with the pseudo-spectral method. a) $\tilde{D} = 0.06$ b) $\tilde{D} = 0.03$.
Figure c) is the density $\rho$ along the white dotted lines shown on figures
a) and b) (red and blue squares respectively). The black curves correspond to
a fit by a sum of Gaussian functions.}
\end{figure}

In two dimensions the behavior of the DK equation is similar,
as no periodic patterns develop for $\alpha \le 2$ or large
$\tD$. Figures~\ref{fig:Patterns_nonoise} a) and b) show that
hexagonal patterns appear for small-enough values of
$\tilde{D}$. Each peak can be reasonably fitted by a
two-dimensional Gaussian (Fig. ~\ref{fig:Patterns_nonoise} c,
which shows one-dimensional cuts of the 2d configurations).

\subsection{Linear stability analysis}
\label{subsec:linearstab}

It is straightforward to check that any constant density
function is a solution of the deterministic Dean equation. We
can thus perform a linear stability analysis around a
homogeneous solution $\rho(\bx)=\rho_0$. In our dimensionless
units all of them are represented by $\rho_0=1$. We consider a
small harmonic perturbation of dimensionless wavenumber
$\bk=\textbf{q} R$, $\rho(\bx,t) = 1 + \delta \rho(\bx,t)$ with
$\delta \rho(\bx,t) = \exp{(\lambda t + i \bk \cdot \bx)}$.
Introducing in (\ref{Dean_nondim}) and linearizing we find the
following growth rate:
\BE
\lambda (k) = -k^2 \left[ \tilde{D} + \hv(k)\right] \ ,
\label{disp_nonoise}
\EE
with $\hv(k)$ the d-dimensional Fourier transform of the
dimensionless interaction potential:
\BE
\hv(\bk) = \int \tv(\bx) e^{- i \bk\cdot \bx} d\bx \ .
\label{Fourier}
\EE
Note that because $\tv(\bx)$ depends only on the modulus $r$ of
$\bx$, $\tv(\bx)=\tv(r)$, the same is valid for $\hv$:
$\hv(\bk)=\hv(k)$, with $k=|\bk|$. Eq.(~\ref{disp_nonoise}) is
the same in any dimension $d$, but the Fourier transform
$\hv(k)$ will be different for different $d$.

\begin{figure}
\includegraphics[width=\columnwidth]{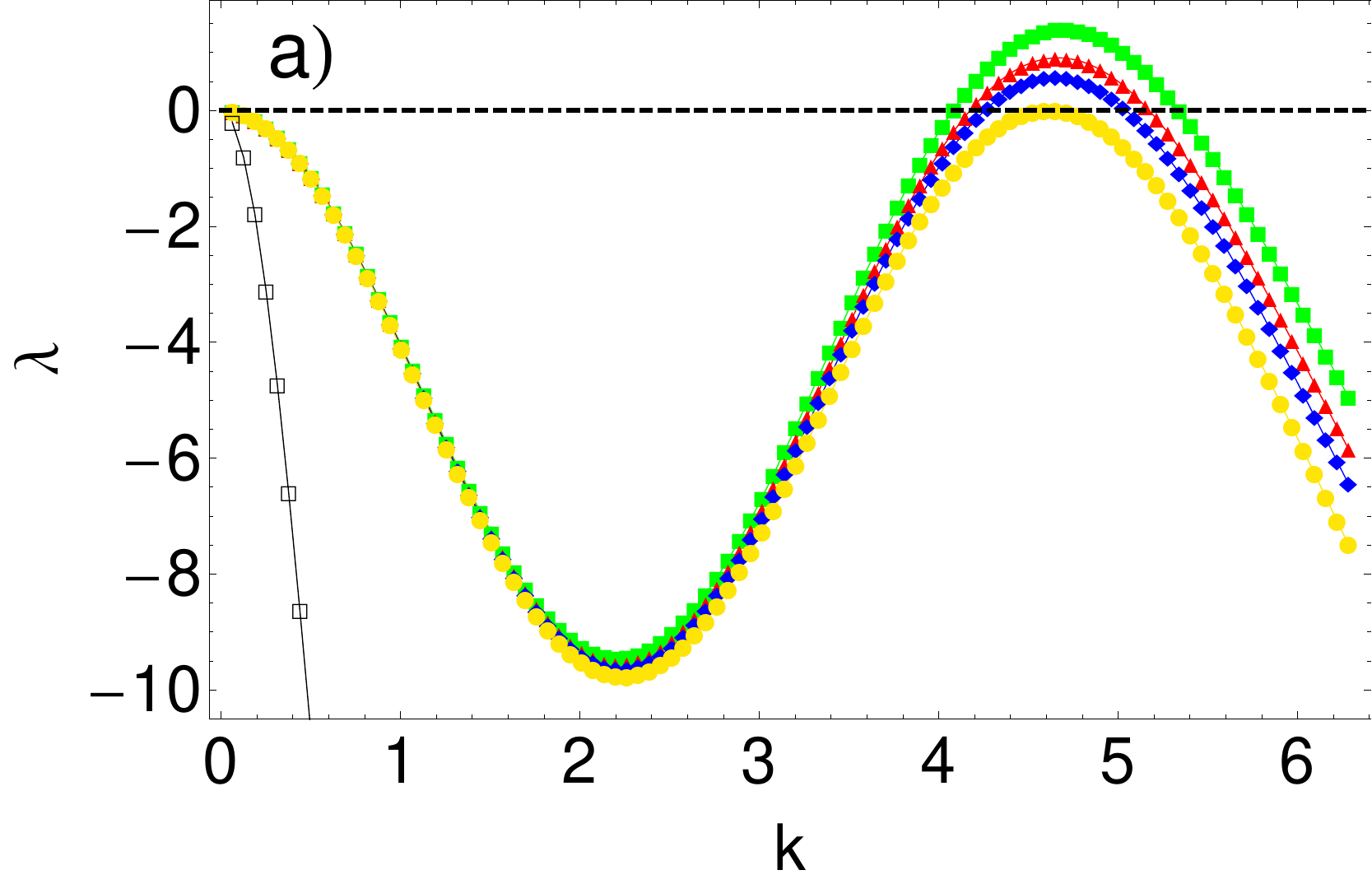}
\includegraphics[width=\columnwidth]{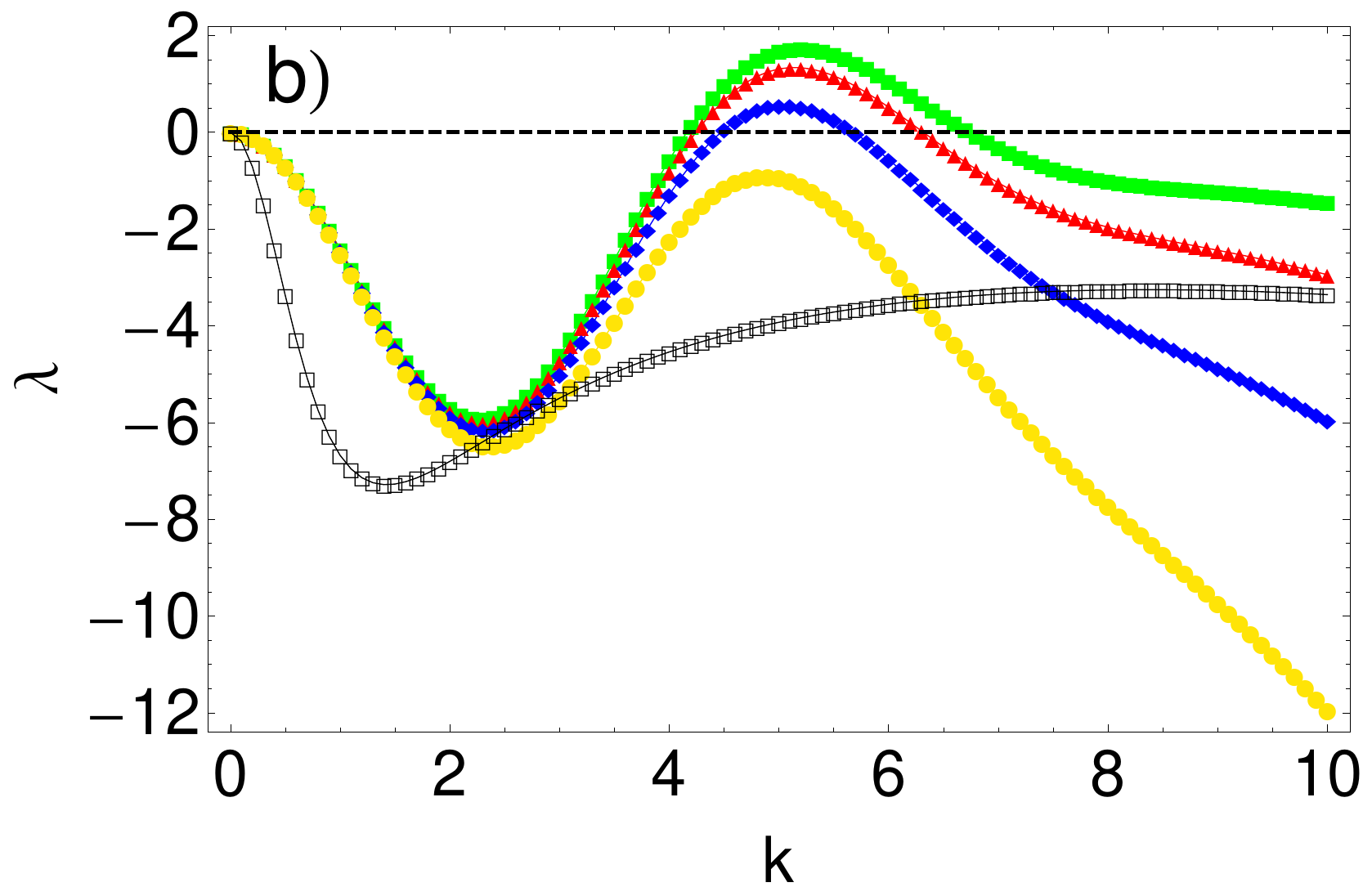}
\caption{\label{fig:stability_nonoise} Growth rates, from Eq.~(\ref{disp_nonoise})
a) 1d. GEM-3 potential with $\tilde{D} = 0.075$ (green squares),
$0.084$ (red triangles), $0.09$ (blue diamonds) and $0.1005$ (yellow disks) and
GEM-1 potential with $\tilde{D} = 0.075$ (black squares). b) 2d.  GEM-3 potential
with $\tilde{D} = 0.015$ (green squares), $0.03$ (red triangles),
$0.06$ (blue diamonds) and $0.12$ (yellow disks) and GEM-1 potential
with $\tilde{D} = 0.015$ (black squares).}
\end{figure}

Figure \ref{fig:stability_nonoise} shows $\lambda(k)$ for
several potentials and parameters, in 1d and 2d.  This growth
rate explains why we never observed patterns with a GEM-1
potential: since its Fourier components are always positive,
the growth rates $\lambda (k)$ are always negative and the
homogeneous state always stable. This can be generalized to any
GEM-$\alpha$ potential with $\alpha <
2$~\cite{Bochner1937,Likos2007}. Also, Eq.(\ref{disp_nonoise})
gives us a precise condition for the onset of patterns when $\hv(k)$ is 
negative in some range of $k$: the
homogeneous state is stable only if we have
\BE
\tD > \tD_c = -\hv_1=|\hv_1| \ ,
\label{stability_criteria}
\EE
where $\hv_1\equiv \hv(k_c)$ and $k_c$ is the wavelength
corresponding to the maximum growth rate. The values of $\hv_1$
are reported in Table~\ref{table1} for several types of
GEM-$\alpha$ potentials. For example with $\alpha=3$, we obtain
$\tD_c \approx 0.1017$ in 1d and $\tD_c \approx 0.0823$ in 2d.
Note that in Fig. ~\ref{fig:structure} c), the transition
threshold seems to be higher with $\tD_c \approx 0.1$. We will
see in section~\ref{subsec:nature} that this is due to the
subcritical nature of the transition. Finally, the intercluster
distance $c=a/R$ always corresponds to a wavelength $k=2\pi/c$
close to $k_c$ as can be seen in Table~\ref{table1} and such as
$\lambda(k) \ge 0$, indicating that the steady-state pattern is
selected by the instability.

\begin{table}
\begin{center}
\textbf{1d}\\
\begin{tabular}{|c|c|c|c|c|c|}
\hline
\hline
$\alpha$ & $\tD_c=-\hv_1$ & $10^2 ~\hv_2$ & $k_c=q_c R$ & $c=a/R$ & $v''(c)$   \\
\hline
3        & 0.1017       & $-0.1108$ &    4.5513     &  1.3805 &  1.7573  \\
4        & 0.1873       & $0.5767$  &    4.5918     &  1.3683 &  2.4787  \\
8        & 0.3326       & $4.3699$  &    4.6519     &  1.3507 &  0.0614  \\
\hline
\hline
\end{tabular}
\\ \vspace{.2cm}
\textbf{2d}\\
\begin{tabular}{|c|c|c|c|c|c|c|}
\hline
\hline
$\alpha$ & $\tD_c=-\hv_1$ & $10^4 ~\hv_2$ & $10^4 ~\hv_{12}$  & $k_c=q_c R$ & $c=a/R$ & $v''(c)$   \\
\hline
3        & 0.0823       & $-5.4727$     & $5.6030$          & 5.0         &  1.4425   &  1.5068  \\
4        & 0.1568       & $-7.4218$     & $139.04$          & 5.1         &  1.3645   &  2.5269  \\
8        & 0.2939       & $-94.321$     & $813.35 $         & 5.2         &  1.2671   &  1.9878  \\
\hline
\hline
\end{tabular}
\end{center}
\caption{Important values of the Fourier transform of some
GEM-$\alpha$ potentials in 1d and 2d. Intercluster distance $a$
is estimated from the location $q_c$ of the maximum growth rate
as $a=2\pi/q_c$, so that $c=2\pi/k_c$} \label{table1}
\end{table}

\subsection{The physical mechanism: effective cluster interactions}
\label{subsec:clustersinteract}

The linear stability analysis provides a clear mathematical
explanation of the instability of the homogeneous state, but it
continues to be counterintuitive to observe clusters composed
of very close particles despite they repel each other. In fact,
for hard spheres the solid state is a crystal of individual
particles, not clusters of them. The reason for cluster
formation is that, despite there is intracluster repulsion, the
particles are also repelled by the particles in the neighboring
clusters. For interactions of the GEM-$\alpha$ type with
$\alpha>2$ the combined repulsion each particle feels from the
ones in neighboring clusters is larger than the repulsion from
the same-cluster particles. We can see this from the following
argument, which also gives us a way to estimate analytically
the cluster width for small $\tD$.

Equation (\ref{Dean_nondim}) can be written as a
particle-conservation equation:
\BE
\partial_t \rho(\bx,t) = -\nabla\cdot {\bf J}(\bx,t) \ ,
\label{conservation}
\EE
with the particle-flux vector given by
\BE
{\bf J}(\bx,t) =-\rho(\bx,t)\int d\bx' \nabla
\tv(\bx-\bx')\rho(\bx',t) - \tD\nabla \rho(\bx,t) \ .
\label{conservationflux}
\EE
We consider zero-flux steady state solutions of
Eq.~(\ref{Dean_nondim}), i.e. solutions with
$\textbf{J}=\textbf{0}$ in (\ref{conservationflux}):
\BE
\rho(\bx)\int d\by\nabla \tilde{v}(\bx-\by)\rho(\by) \nonumber
= - \tilde{D}\nabla \rho(\bx),
\EE
which, after integration gives us
\BE
\rho(\bx) = \exp \left[ \frac{1}{\tD}\left(\mu - \int d\by
\,\rho(\by) \tilde{v}(\bx-\by) \right)\right]. \label{rho_eq}
\EE
$\mu$ is an integration constant, to be fixed by the
normalization of $\rho$, and that can be identified with a
chemical potential.  We consider Eq.~(\ref{rho_eq}) as an
iterative procedure to obtain the steady configuration: by
substituting in the right-hand-side of the equation a first
approximation to the density, the left-hand-side will give an
improved one. In the limit of $\tD \to 0$, for which the
density becomes a periodic arrangement of narrow clusters, a
sensible first approximation would be an array of
delta-function clusters.  In the one-dimensional case this is
\BE
\rho(y) \approx N_p\sum_n\delta(y-n c) \ . \label{sumofdeltas}
\EE
The sum over $n$ is over all clusters in the system, and the
origin of coordinates is such that there is a cluster at $y=0$.
$c$ is the intercluster distance. It would be close to the most
unstable wavelength, i.e. $c\approx 2\pi/k_c$. $N_p$ is the
number of particles contained in each cluster. For identical
clusters it is $N_p=N c /L=c$, where the last inequality arises
since we are using units such that $\rho_0=R=1$.

\begin{figure}
\includegraphics[width=\columnwidth]{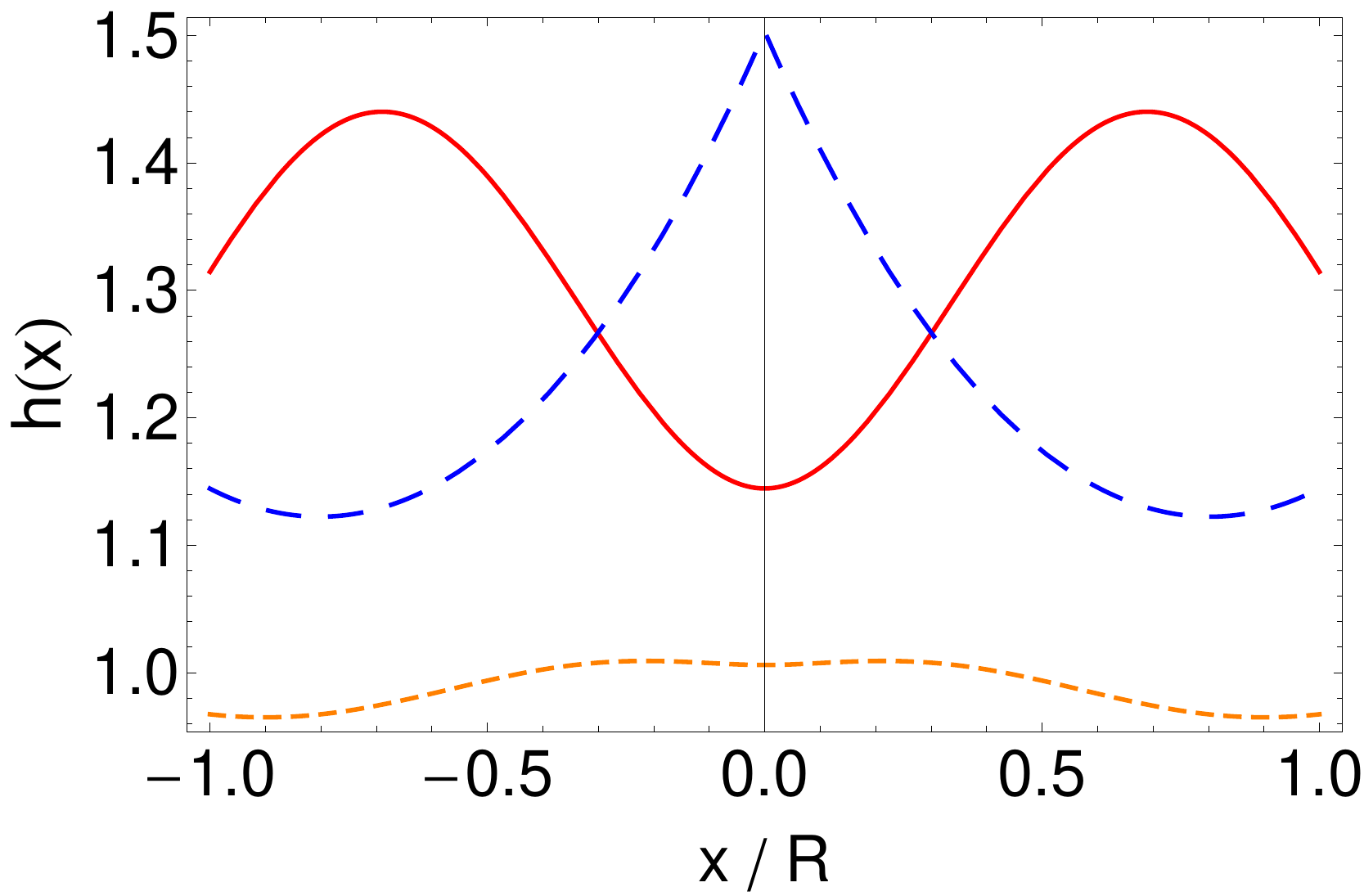}
\caption{\label{fig:EffectivePot}Effective potential $h(x)$. Solid:
GEM-3 potential, with intercluster distance $c=a/R = 1.38$, which is
(see Table \ref{table1}) the periodicity given by the linear stability analysis.
Dashed: $h(x)$ for GEM-1. We use the same value $c=1.38$, although
the peak at the origin is independent of the intercluster separation.
Dotted: $h(x)$ for the GEM-3 potential but for a larger intercluster separation $c=1.8$.}
\end{figure}

Inserting this first approximation back into Eq.~(\ref{rho_eq})
we obtain the improved approximation:
\BE
\rho(x) \approx \exp \left[ \frac{1}{\tD}\left(\mu - N_p\sum_n
\tilde{v}(x-n c) \right)\right]. \label{rho_eq1d}
\EE
The GEM-$\alpha$ functions $\tv(x)$ are rapidly decreasing
towards zero as soon as $x>R=1$. Then, at each particular
location $x$, it is a good approximation to take into account
only the contributions to the sum from the closest cluster and
from their two neighbors, provided cluster separation is larger
than $R$. For example, if we want to approximate the density
close to $x\approx 0$ we can consider the terms with $n=0$
and $n=\pm 1$ in (\ref{rho_eq1d}):
\BE
\rho(x\approx 0) \approx e^{\frac{\mu}{\tD}} \exp\left(-
\frac{N_p}{\tD} h(x) \right)\ , \label{rho_3clust}
\EE
where
\BE
h(x)\equiv  \tv(x) +\tv(x+c) +\tv(x-c) \ .
\label{hpotential}
\EE
The shape of the cluster close to the origin is determined by
the function $h(x)$ which acts as an {\sl effective potential}
felt by a test particle at position $x$. $h(x)$ combines the
repulsion from the particles in the cluster at the origin,
$\tv(x)$, with the repulsion from the particles in the
neighboring ones $\tv(x\pm c)$. If the resulting $h(x)$ has a
minimum at the origin our assumption of a narrow cluster there
would be justified. On the contrary, a maximum of $h(x)$ at the
origin implies that the internal repulsion dominates, our
approximations (\ref{sumofdeltas}) and (\ref{rho_eq1d}) would
be inadequate, and the iteration procedure does not converge to
a steady configuration consisting  on well separated clusters.
Figure~\ref{fig:EffectivePot} shows the function $h(x)$ for
various GEM-$\alpha$ potentials. A maximum at the origin and
then lack of convergence to localized clusters occurs when
$\alpha<2$ (see Fig. ~\ref{fig:EffectivePot} for GEM-1). A
change of behavior at the origin occurs precisely at
$\alpha=2$. This can be seen by expanding $h(x)$ close to the
origin: Using $\tv(x)=\exp(-|x|^\alpha)\approx 1-|x|^\alpha
+\ldots$ and that $\tv(x)$ is analytic at any $x\neq0$ we find
\BE
h(x)\approx 1+ 2 \tv(c) + \tv''(c) x^2 -|x|^\alpha + \ldots
\label{hexpan}
\EE
If $\alpha<2$ the dominant term is $-|x|^\alpha$ which gives a
maximum at the origin and then Eq.~(\ref{rho_eq1d}) does not
give localized clusters at the intended positions. When
$\alpha>2$ the quadratic term dominates at small $x$ and the
maximum or minimum character of $h(x)$ at the origin is
determined by the second derivative or curvature $\tv''(c)$,
which for the GEM-$\alpha$ potential is:
\BE
\tv''(c) = c^{\alpha-2}\alpha (1-\alpha+\alpha c^\alpha)
e^{-c^\alpha} \ .
\EE
For very small intercluster separation $c$, $v''(c)$ becomes
negative and then the situation is similar to $\alpha<2$. But
in the tail of the potential, i.e. if $c\gtrsim 1$ (as when
cluster distance is given by the linear instability, see Table
\ref{table1}) this curvature is always positive and the
iterative procedure will converge (if $\tD$ is small) to a
steady solution made of localized clusters. $h(x)$ for GEM-3
and intercluster distance $c$ given by the linear stability
analysis is plotted in Fig. ~\ref{fig:EffectivePot}, showing a
clear confining character at the origin. Values of the
linearly-determined $c$ and of $v''(c)$ for other values of
$\alpha$ are in Table \ref{table1}. When the intercluster
distance is too large, however, the influence of the
neighboring clusters becomes weaker. As shown in Fig.
~\ref{fig:EffectivePot} for a large $c=1.8$, the minimum
character of the origin is preserved, but the minimum is very
shallow and the absolute minima are not there, but at lateral
positions. Thus, the iterative procedure starting with the
ansatz Eq.~(\ref{sumofdeltas}) will not converge to a proper
steady solution neither in this case. This gives a range of
periodicities (roughly $R \lesssim a \lesssim 2R$) for which
intercluster repulsion under GEM-$\alpha$ potentials with
$\alpha>2$ lead to localized clusters despite the internal
repulsion existing in all of them.

We have focused in this section on the one-dimensional case,
but it is easy to see that the general ideas and mechanisms are
valid in any dimension $d$ and in fact we use them in the next
subsection to obtain quantitative expressions for cluster width
and shape in 1d and in 2d.

\begin{figure}
\includegraphics[width=\columnwidth]{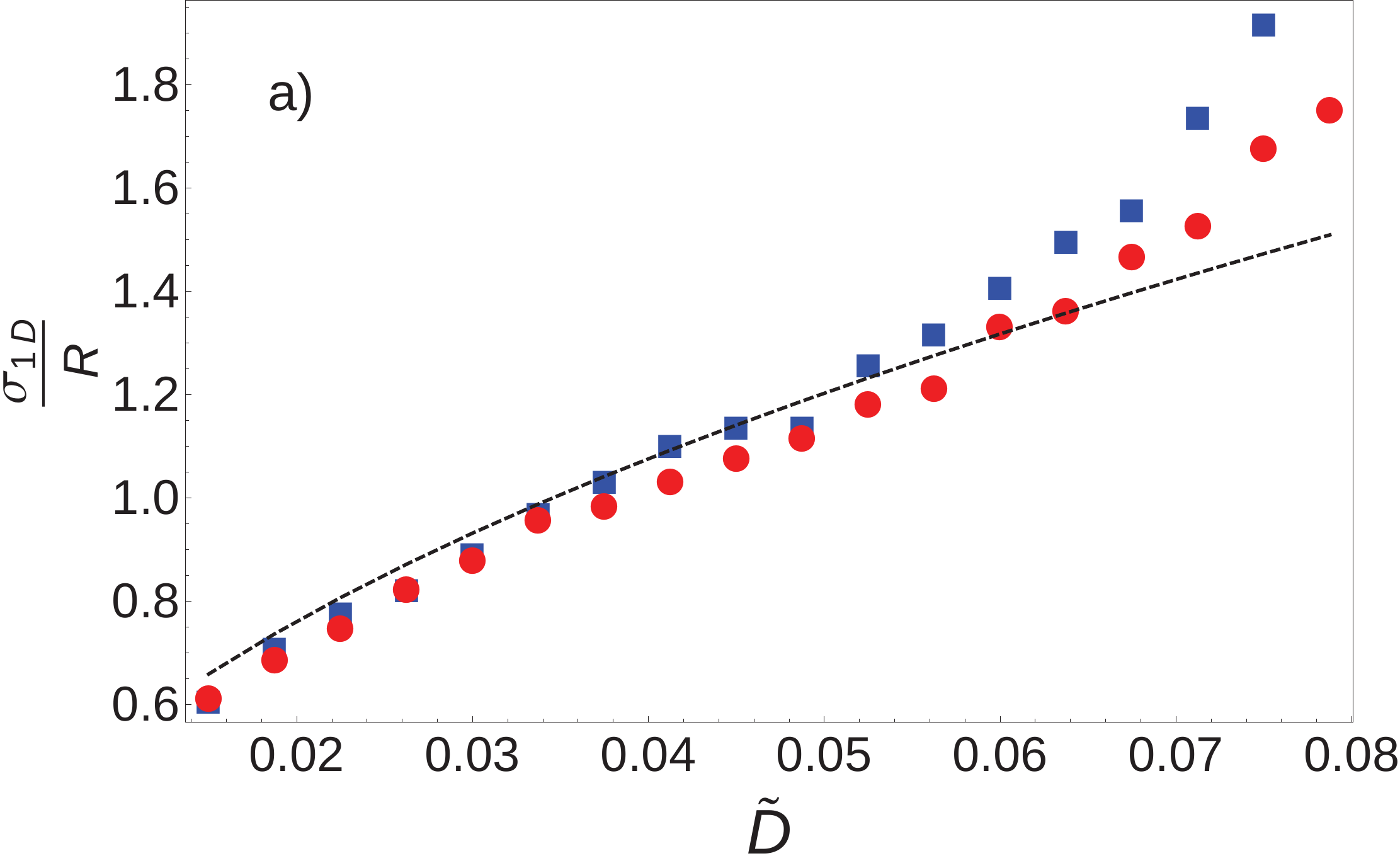}
\includegraphics[width=\columnwidth]{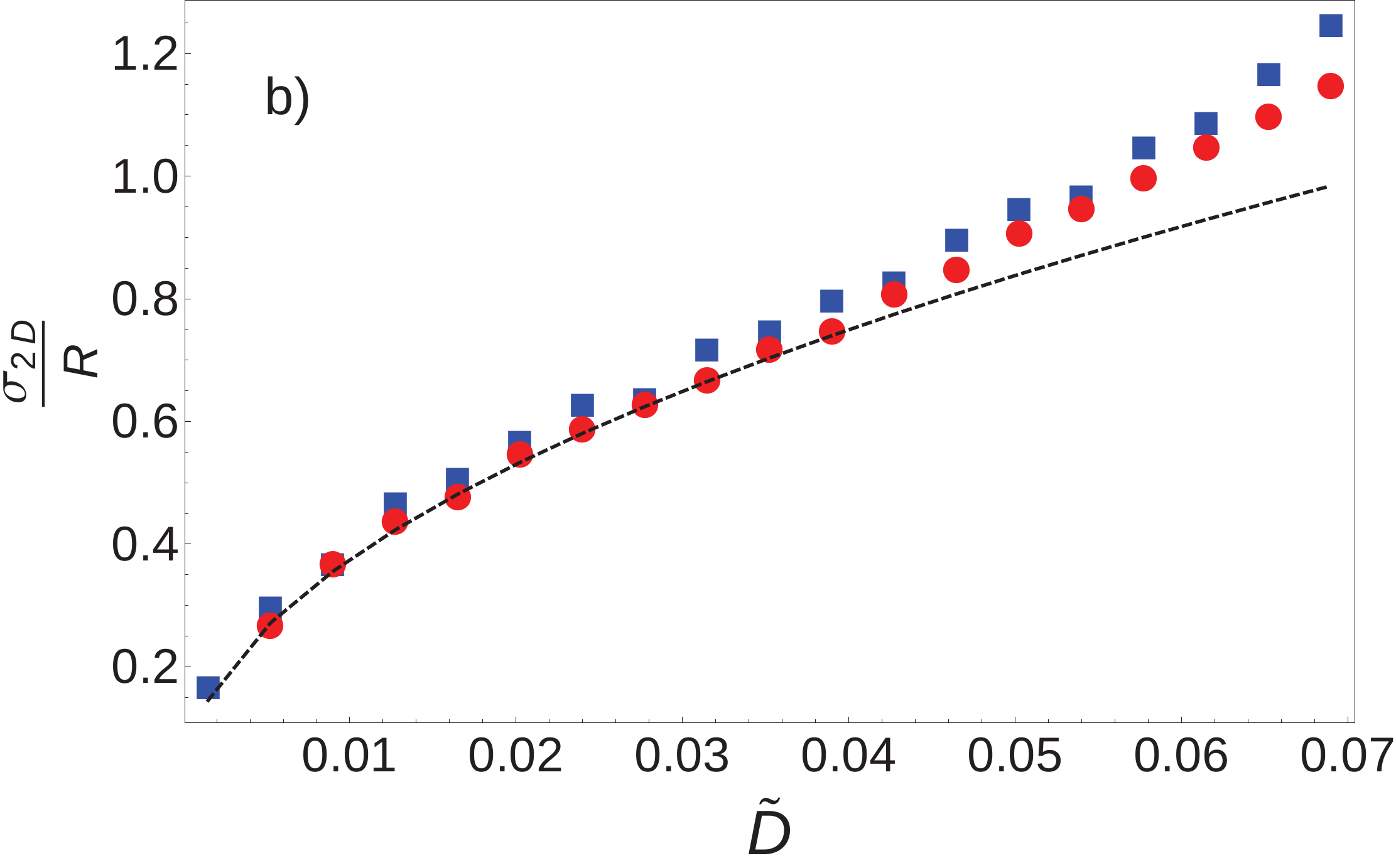}
\caption{\label{fig:sigma1d} Clusters width as a function of $\tilde{D}$ for a GEM-3 potential.
Red circles: from the steady density solution of the DK equation. Blue squares: from a coarse-grained
density in particle simulations.
In both cases, the Gaussian fit was made focusing on the top of the clusters.
a) One-dimensional case. Dashed black line is Eq.~(\ref{gaussian_width_1Dbis}).
b) Two-dimensional case. Dashed black line is Eq.~(\ref{gaussian_width_2D}).
The intercluster distance $c$ used in the analytical formula was obtained
from the position of the second peak of the radial distribution
function $g^{(2)}(r)$ (see Fig. ~\ref{fig:structure} a)), which gives a result
slightly better than using the linear instability value $2\pi/k_c$. In the particle case the
coarse graining in 1d and in 2d was done as in Figs. \ref{fig:1Dparticles} and \ref{fig:P_simu}
respectively.}
\end{figure}

\subsection{Shape and width of clusters}
\label{subsec:clustershape}

The above arguments shed additional insight on the results of
the linear stability analysis: homogeneous distributions become
unstable against pattern formation of periodicity precisely in
the range which allows the particles to remain confined in
well-localized clusters. All the explanations rely on having
narrow clusters, which is only justified if $\tD$ remains
sufficiently small. In this limit (and for $\alpha>2$),
expressions (\ref{rho_3clust})-(\ref{hexpan}), valid for
$x\approx0$, can be used to provide an estimation of the
central cluster shape and width. In one dimension we have
\BE
\rho(x) \approx e^{\frac{\mu - N_p(1+2\tv(c))}{\tD}}
e^{-\frac{N_p}{\tD} \tv''(c) x^2} \equiv
\frac{N_p}{\sqrt{2\pi\sigma^2}} e^{-\frac{x^2}{2\sigma^2}} \ ,
\EE
so that each cluster can be approximated by a Gaussian of width
\BE
\sigma=\sqrt{\frac{\tD}{2 N_p \tv''(c)}} \ .
\label{gaussian_width_1D}
\EE
Denoting $\sigma_{1D}$ this width in the original unscaled
units, and using that $N_p=c=a/R$ this expression can be
written as:
\BE
\frac{\sigma_{1D}}{R}=\sqrt{\frac{D}{2\epsilon\rho_0 a
\tv''(c)}} \label{gaussian_width_1Dbis}
\EE

Figure~\ref{fig:sigma1d} a) compares this formula with the
width obtained from 1d numerical simulations of the particle
system and of the DK equation. We see that for the lowest
values of $\tD$, the agreement is really good in both cases.
For higher values of $\tD$, our calculations tend to
underestimate the width of the clusters. This is coherent with
our hypothesis as we assumed very small values of $\tD$ and
well separated clusters.

In this $\tD\rightarrow 0$ limit the full density, according to
Eq.~(\ref{rho_eq1d}) is an array of Gaussian peaks:
\BE
\rho(x)\approx  \frac{N_p}{\sqrt{2\pi\sigma^2}} \sum_n
e^{-\frac{(x-n c)^2}{2\sigma^2}} \ .
\EE
The maximum (peak) value of the density is then (using $N_p=c$)
\BE
\rho_{max}=\frac{1}{\sqrt{2\pi}}\frac{c}{\sigma} \ ,
\EE
or in terms of unscaled variables (and $a= c R$):
\BE
\frac{\rho_{max}}{\rho_0}=\frac{1}{\sqrt{2\pi}}
\frac{a}{\sigma_{1D}} \ .
\label{rhomax1d}
\EE
This expression is plotted in Fig. ~\ref{fig:biff1d} as a
dashed line, and compared with the numerically obtained 1d
steady solution of the DK equation and with particle
simulations. As expected, it becomes accurate when
$\tD\rightarrow 0$ but becomes increasingly worse when $\tD$
approaches the transition point. This opposite regime will be
discussed in the next subsection.

All the previous calculations can be essentially repeated in
2d. The only major difference is the starting point as we now
have to replace Eq.~(\ref{sumofdeltas}) by a hexagonal lattice
of delta functions:
\BE
\rho(\by) \approx N_p \sum_{n}
\sum_{m} \delta (\by+n \boldsymbol{c}_1 + m \boldsymbol{c}_2),
\label{sumofdeltas2d}
\EE
where $\boldsymbol{c}_1$ and $\boldsymbol{c}_2$ are two vectors
of norm $c$ necessary to generate the hexagonal patterns.
Following the same method as in one dimension, we find that the
clusters have once again a Gaussian shape with width $\sigma$
given by
\BE
\sigma=\sqrt{\frac{2\tD}{3\sqrt{3} \alpha^2 c^\alpha
(c^\alpha-1)e^{-c^\alpha}}} \ ,\label{gaussian_width_2D}
\EE
which gives us in unscaled units
\BE
\frac{\sigma_{2D}}{R}=\sqrt{\frac{2 D}{\epsilon \rho_0 R^2
3\sqrt{3} \alpha^2 c^\alpha (c^\alpha-1)e^{-c^\alpha}}} \ .
\label{gaussian_width_2D}
\EE
This formula is compared with numerical measurements in Fig.
~\ref{fig:sigma1d} b) for $\alpha=3$. The agreement with
particle and DK data is once again very good as long as the
value of $\tilde{D}$ is small enough.

\subsection{The neighborhood of the transition point}
\label{subsec:nature}

In the previous subsection we obtained an accurate description
of the patterns formed at small $\tD$. At the same time the
theoretical arguments to obtain it gave useful insight into the
mechanisms of the cluster crystal formation. But this
description became inaccurate as $\tD$ increased, and can not
describe the neighborhood of the instability point. Here we
focus on that regime, using weakly nonlinear expansions
\cite{Walgraef2012} close to the bifurcation point $\tD_c =
-\hv_1 = |\hv_1|$. Although this type of description is
appropriate for solutions of the deterministic DK equation, it
should be recognized that fluctuations effects tend to be
noticeable close to instability points, and then it is not
guaranteed that the expressions obtained in the present
subsection would be accurate for the stochastic particle
system.

\begin{figure}
\includegraphics[width=\columnwidth]{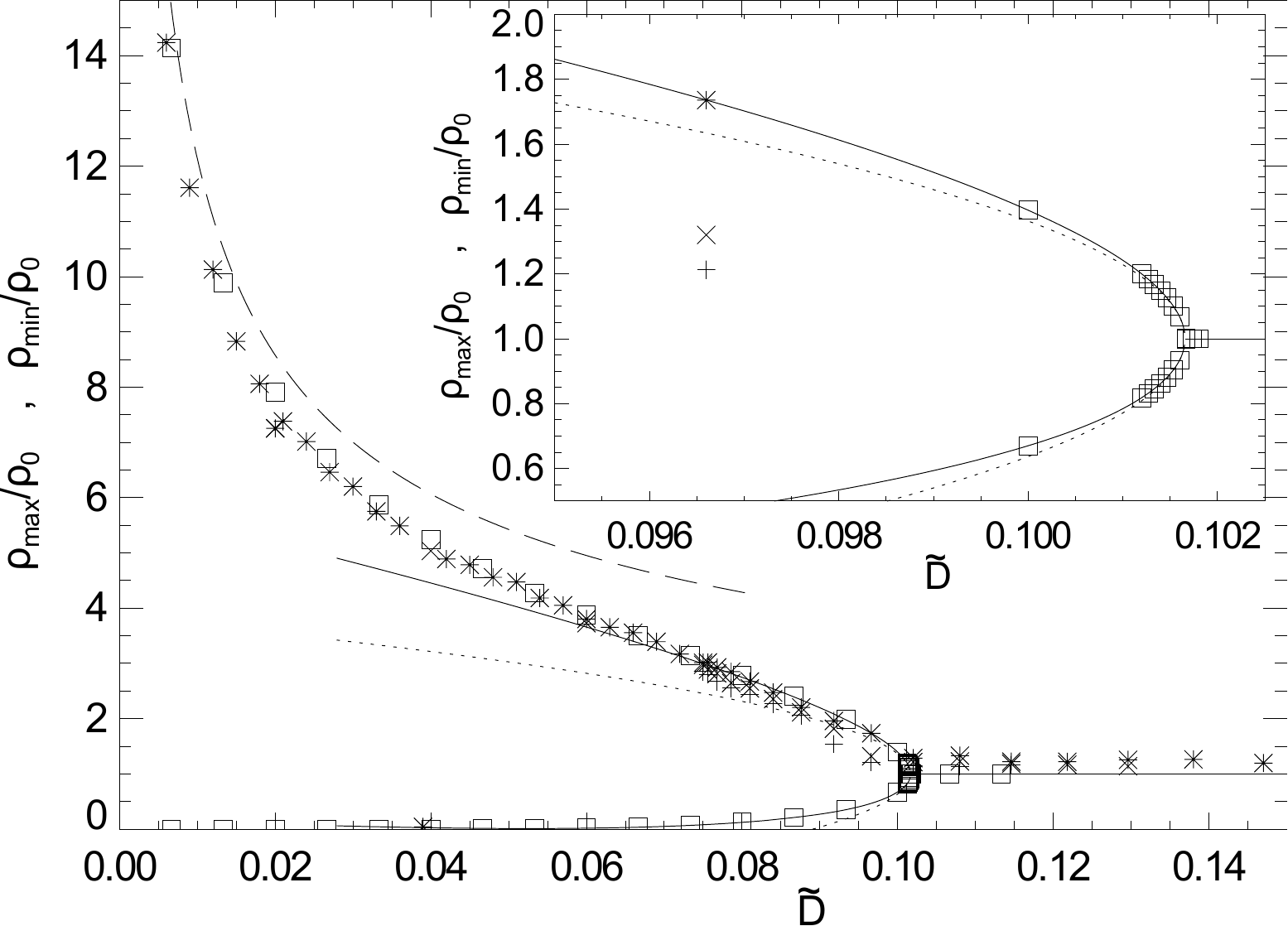}
\caption{\label{fig:biff1d} For each value of $\tD$ the maximum and the minimum value of the steady
density in 1d for $\alpha=3$ are plotted. The inset gives an enlargement of the bifurcation region.
Squares ($\Box$): values from direct numerical simulation of the DK equation. *, $\times$ and +:
values from a coarse-grained density in particle simulations at $\rho_0=1000$, $1500$ and
$2000$ respectively ($R=0.1$, $\epsilon=0.0333$).
Solid line: the weakly nonlinear approximation Eq.~(\ref{amplitude1d}),
including up to the second harmonic. Dotted line: only the first harmonic term
in Eq.~(\ref{amplitude1d}). Dashed line: the maximum of the pattern as given by the
Gaussian approximation, Eq.~(\ref{rhomax1d}). In the particle case the
coarse graining was done as in Fig. (\ref{fig:1Dparticles}).
}
\end{figure}

We start with the 1d case. Details of the calculations are in
the Appendix. We use a weakly nonlinear expansion in powers of
a small parameter which turns out to be the square root of the
distance to the bifurcation point, $\sqrt{|\hv_1|-\tD}$. We
obtain an amplitude equation describing the dynamics close to
the bifurcation. From it, the steady periodic solution up to
second order in the small parameter reads
\BE
\rho_{st}(x)=1+K (|\hv_1|-\tD)^{1/2} \cos(k_c x)+\frac{2
(|\hv_1|-\tD)}{2\hv_2+|\hv_1|} \cos(2 k_c x) \ ,
\label{amplitude1d}
\EE
where $\hv_2=\hv(2k_c)$,
\BE
K \equiv 2\left(
\frac{2(1+\hv_2/|\hv_1|)}{2\hv_2+|\hv_1|}\right)^{1/2} \ ,
\label{Kamplitude1d}
\EE
and the pattern position has been chosen to be such that there
is a maximum at $x=0$. Values of $\hv_1$ and $\hv_2$ for
different values of $\alpha>2$ are in Table~\ref{table1}.
Figure~\ref{fig:biff1d} compares this expression with maximum
and minimum values of the density distribution for $\alpha=3$,
numerically obtained from the 1d DK equation and from particle
simulations. It is seen that the approximation including only
the first harmonic is accurate just very close to the
instability point and that the range of validity is improved by
including the second-order term. Note that the continuous
character of the bifurcation in the DK equation is correctly
predicted by the theoretical formula. For small values of $\tD$
the agreement becomes rather poor, as expected, but the maximum
values of the density can then be predicted by the Gaussian
approximation and Eq.~(\ref{rhomax1d}). It is also worth
mentioning that, in most of the $\tD$ range, proper scaling is
observed when plotting the maximum and minimum values of the
coarse-graining particle density in terms of the dimensionless
quantities identified from the deterministic DK equation, and
that only small differences between the DK density and the one
obtained from particle simulations are observed. This confirms
that neglecting noise, which has allowed us to get analytic
insight, is a good approximation in most of the parameter
range. The exception is the neighborhood of the bifurcation
point of the DK equation since, as commented before, we do not
expect a sharp transition for the particle system in 1d. The
effect of this somehow smoother transition is an apparent
shifting of the critical point to smaller values of $\tD$ for
particle simulations, an expected result of the presence of
fluctuations. This also explains why the local density was more
sinusoidal on the top right part of Fig. ~\ref{fig:1Dparticles}
than on Fig. ~\ref{fig:Patterns_nonoise_1d} b): as the particle
simulations are noisy, at this value of $\tD$ the system is
deeper into the periodic state and further away from the
transition point in the DK case.

\begin{figure}
\includegraphics[width=\columnwidth]{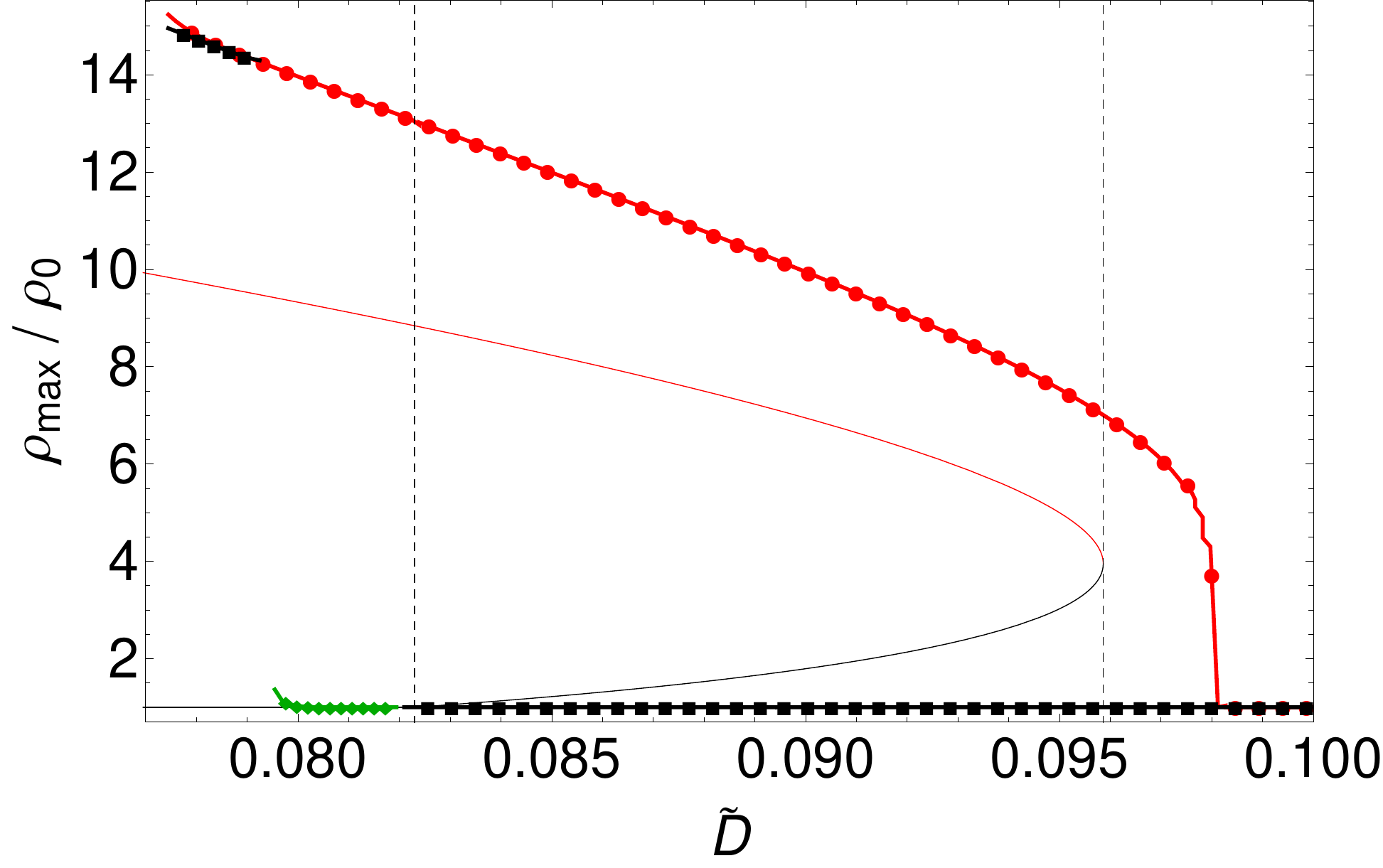}
\caption{\label{fig:hysteresis} Maximum value $\rho_{max} / \rho_0$ of the steady density solution
of the DK equation in 2d for $\alpha=3$. We start from $\tD =0.077$, slowly increase it up to $\tD =0.102$ (red disks)
and then slowly decrease $\tD$ back to its initial value
(black squares and green diamonds). The plot clearly highlights a
hysteretical behavior. The green diamonds indicate that  $\rho_{max}/\rho_0$
remains close to $1$ during the times accessible to our simulation, but
clearly increasing, even if very slowly. For the black squares in the lower branch $\rho_{max}/\rho_0$
is observed to decrease towards 1. Thus the behavior is consistent with the theoretical threshold
for stability of the homogeneous solution ($\tD\approx 0.0823$, see Table \ref{table1},
marked with a dotted vertical line) and we expect the green diamonds to reach the upper density branch
in a sufficiently long simulation. The thin undotted lines give the
theoretical prediction for $\rho_{max}$ from Eq. (\ref{Kamplitude2d}): two branches, giving
the upper one a prediction for the amplitude of the stable hexagonal density. The
right dotted vertical line indicates its turning point.}
\end{figure}

We now turn out to two-dimensional systems. Following a similar
weakly nonlinear procedure, we find the following expression
for the density for $\tD$ close to $\tD_c$ (see details in the
Appendix):
\begin{widetext}
\BE
\rho_{st}(\bx)=1+2\delta_0\left( \cos (\bk_1\cdot\bx) +
\textrm{cyclic} \right)  + \frac{\delta_0^2
|\hv_1|}{(|\hv_1|+\hv_2)}
 \left( \cos( 2 \bk_{1}\cdot\bx) + \textrm{cyclic}  \right) + \frac{2 \delta_0^2
|\hv_1|}{|\hv_1|+\hv_{12}} \left( \cos( \bk_{12}\cdot\bx) +
\textrm{cyclic}  \right) \ , \label{Kamplitude2d}
\EE
\end{widetext}
where $\delta_0$ is, up to second order, a positive solution of
\BE
\frac{D-|\hv_1|}{|\hv_1|}=\delta_0 - \delta_0^2 \left(
\frac{|\hv_1|+2\hv_2}{2(|\hv_1|+\hv_2)} +
\frac{|\hv_1|+3\hv_{12}}{|\hv_1|+\hv_{12}} \right) \ .
\label{Kamplitude2d_delta}
\EE
The terms in `cyclic' are cosines similar to the ones
explicitly shown but with arguments obtained by changing
cyclically ($1\rightarrow 2 \rightarrow 3 \rightarrow 1$) the
subindices of vectors $\bk_1$ and $\bk_{12}$ (which are defined
in the Appendix). In this way $\rho_{st}(\bx)$ has hexagonal
symmetry, with periodicity determined by $k_c=|\bk_1|$.
Figure~\ref{fig:hysteresis} shows the behavior with respect to
$\tD$ of the maximum value of $\rho_{st}/\rho_0$ according to
Eqs.~(\ref{Kamplitude2d}) and (\ref{Kamplitude2d_delta}) and
compares them with the numerical results of the 2d DK equation.
Equation (\ref{Kamplitude2d_delta}), quadratic in $\delta_0$,
gives two different branches for $\delta_0$ in a range of
$\tD$, coalescing and disappearing at a turning point. Only the
upper branch is stable. Then, the bifurcation is subcritical in
2d, with two stable steady densities existing near the critical
point, the homogeneous $\rho=\rho_0$ and the hexagonal density
with amplitude determined by the upper-branch solution of
(\ref{Kamplitude2d_delta}). In agreement with the analytic
approach, a hysteretic behavior is clearly visible in the
DK simulations. The upper branch corresponds to the hexagonal
pattern that discontinuously becomes homogeneous when
increasing $\tD$ beyond $\tD\approx 0.0957$. The quantitative
agreement between simulations and theory is rather poor as
Eq.~(\ref{Kamplitude2d}) systematically underestimates the
density peaks. This is a consequence of the theory being an
expansion in the neighborhood of the point $(\tD=\tD_c,
\rho/\rho_0=1)$, whereas the interesting upper branch of the
density its quite far from there. For smaller values of $\tD$
the 2d Gaussian approach gives a better description, as
confirmed by Fig. \ref{fig:sigma1d}b. We note however that the
turning point predicted from Eqs. (\ref{Kamplitude2d}) and
(\ref{Kamplitude2d_delta}) gives a reasonable approximation to
the numerical location of the jump to homogeneous density in
the DK case. Although it is out of the scope of the present
paper to elucidate if the sharp jump in the maximum of the
structure factor in the particle system (see Fig.
\ref{fig:structure} c) is actually continuous or discontinuous,
indicating a continuous or discontinuous melting, we note that
the discontinuous jump occurring in the deterministic DK steady
solution gives a good approximation ($\tD\approx 0.0957$, a
value of $\tD$ larger than the linear $\tD_c=|\hv_1|$) for the
location of the particle transition, as seen in Fig.
\ref{fig:structure} c.

\section{Dynamics with soft attractive interactions}
\label{sec:attractive}

For completeness, we study in this section the case {\it
opposite} to the previous repulsive situation, i.e. particles
interacting via purely attractive forces. The relevance of this
problem is reflected in many fields of physics or biology
dealing with the problem of particles attracting each other
\cite{Topaz2006,Chavanis2008a}.

\begin{figure}
\includegraphics[width=\columnwidth]{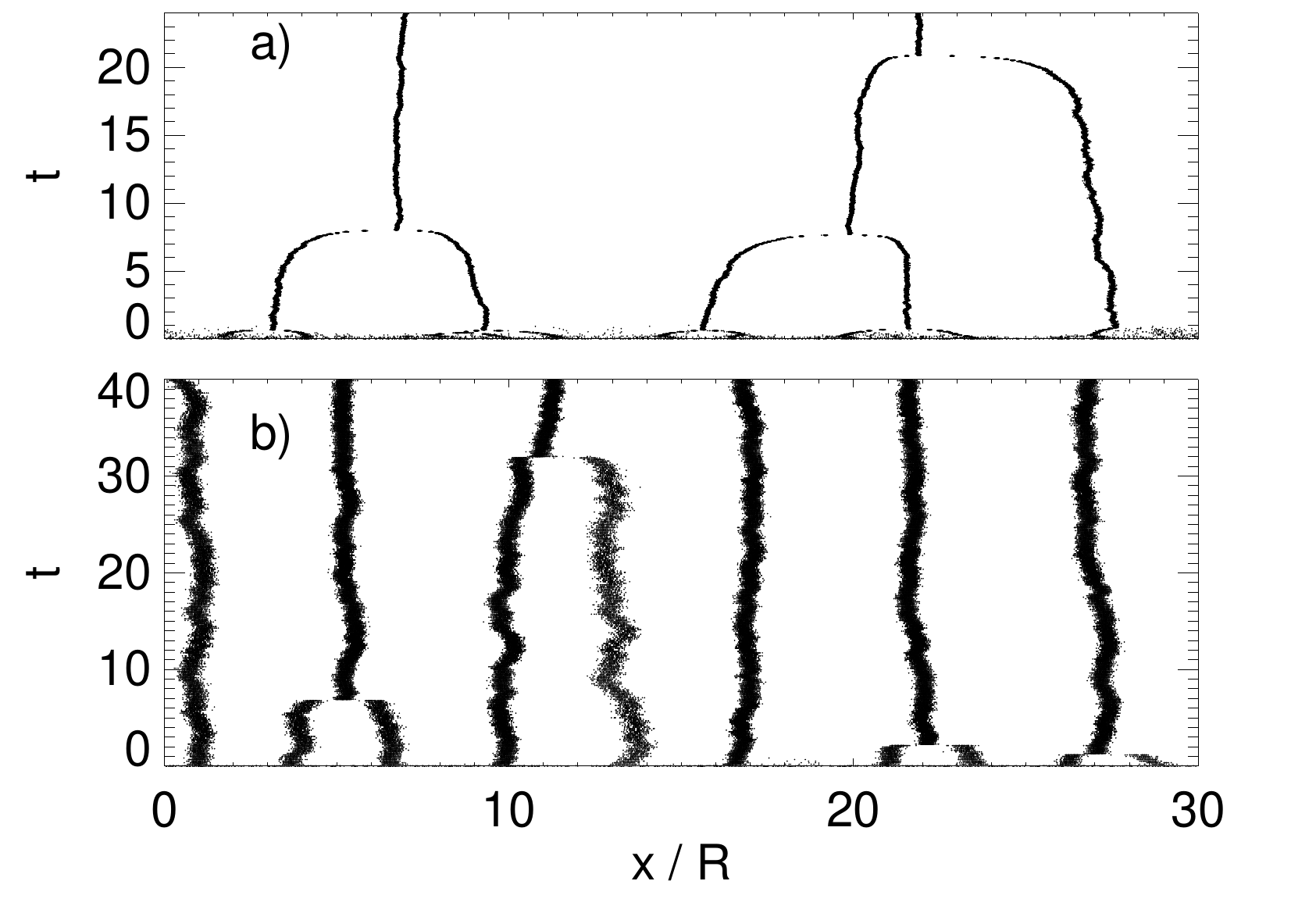}
\caption{\label{fig:particles1d_attractive} Dynamics of 600 Brownian particles
with attractive interactions in a periodic domain of size $L=3$, so that $\rho_0=200$.
$\epsilon=-0.33$, $R=0.1$, and $D=3.96\times 10^{-3}$, so that $\tD=6\times 10^{-4}$. a) GEM-1 attractive potential.
b) GEM-3 attractive potential.}
\end{figure}

\begin{figure}
\includegraphics[width=\columnwidth]{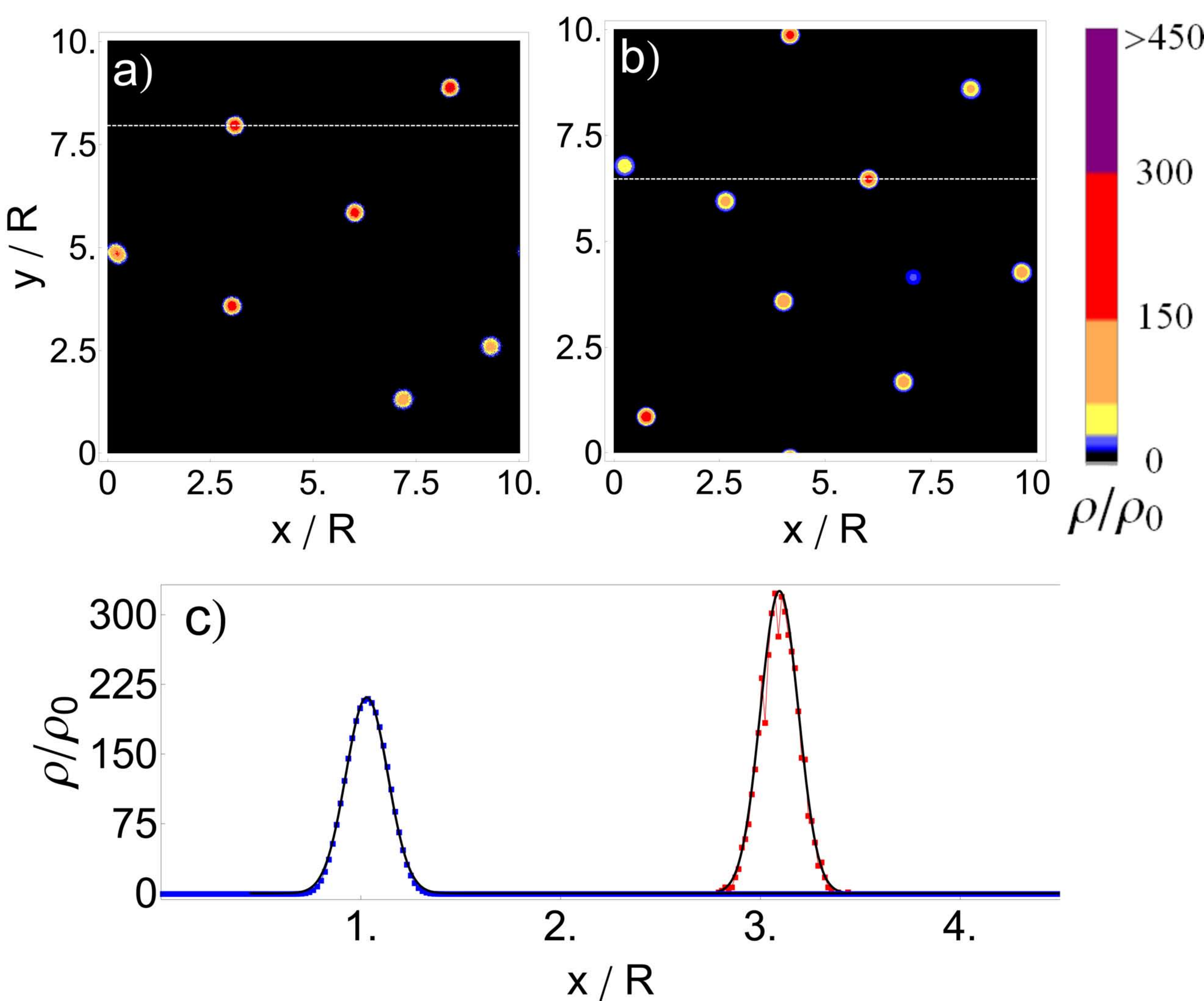}
\caption{\label{fig:patterns_attractive} Density functions at large times
for the attractive GEM-3 potential and $\tilde{D} = 0.0975$.
a) Coarse-grained density
of particle simulations. $R=0.1$, $\epsilon = 0.0333$, $N=2000$, $L=1$ (so that $\rho_0 = 2000$).
The coarse graining procedure is as in Fig. \ref{fig:P_simu}. b) Integration
of Eq.~(\ref{Dean_nondim}) with the pseudo-spectral method.
Figure c) is the density $\rho / \rho_0$ along the white dashed lines shown on
Figs. a) and b) (red and blue squares respectively), shifted for convenience.
The black curves correspond to a fit by a sum of Gaussian functions.}
\end{figure}

We consider Brownian dynamics as in Eq.~(\ref{Brownian}) with a
potential given also by Eq.~(\ref{GEM_def}), but now it is
attractive, so that $\epsilon <0$. The dimensionless version of
the potential is $\tv(\bx)=-\exp(-|\bx|^\alpha)$. In Fig.~\ref{fig:particles1d_attractive} we show a spatiotemporal plot
of the 1d particle positions for both $\alpha =1$ and $\alpha =
3$, starting from random initial conditions. Despite the
visible differences between the two cases, the qualitative
features of the dynamics are the same: in both situations
clusters periodically spaced emerge at short times. Then
clusters attract each other and coalesce. In this coarsening
process the pattern periodicity increases, although at late
times cluster separation becomes progressively more irregular
and each cluster behaves essentially as isolated. As in the
repulsive case, if $\tD$ is sufficiently large, cluster
formation does not occur. The same phenomenology is observed in
2d, and this is also the behavior of the solutions of the 1d
and 2d DK equation with attractive potential. Figure
~\ref{fig:patterns_attractive} displays 2d late time
configurations for both the particle dynamics and the DK
description. The figure shows also that the shape of the
clusters is approximately Gaussian.

Because of the attraction one would expect a collapse of all
the particles in a single cluster. In fact, this is what
happens but at extremely long times. The clustering is a
consequence of particle's attraction, and several aggregates
remain if the attraction is weak and the clusters are distant
enough from each other. In the case of the noiseless DK equation, if the interaction has a
strictly finite range, clusters located farther apart than this
range do not coalesce and the pattern would remain stationary.

\begin{figure}
\includegraphics[width=\columnwidth]{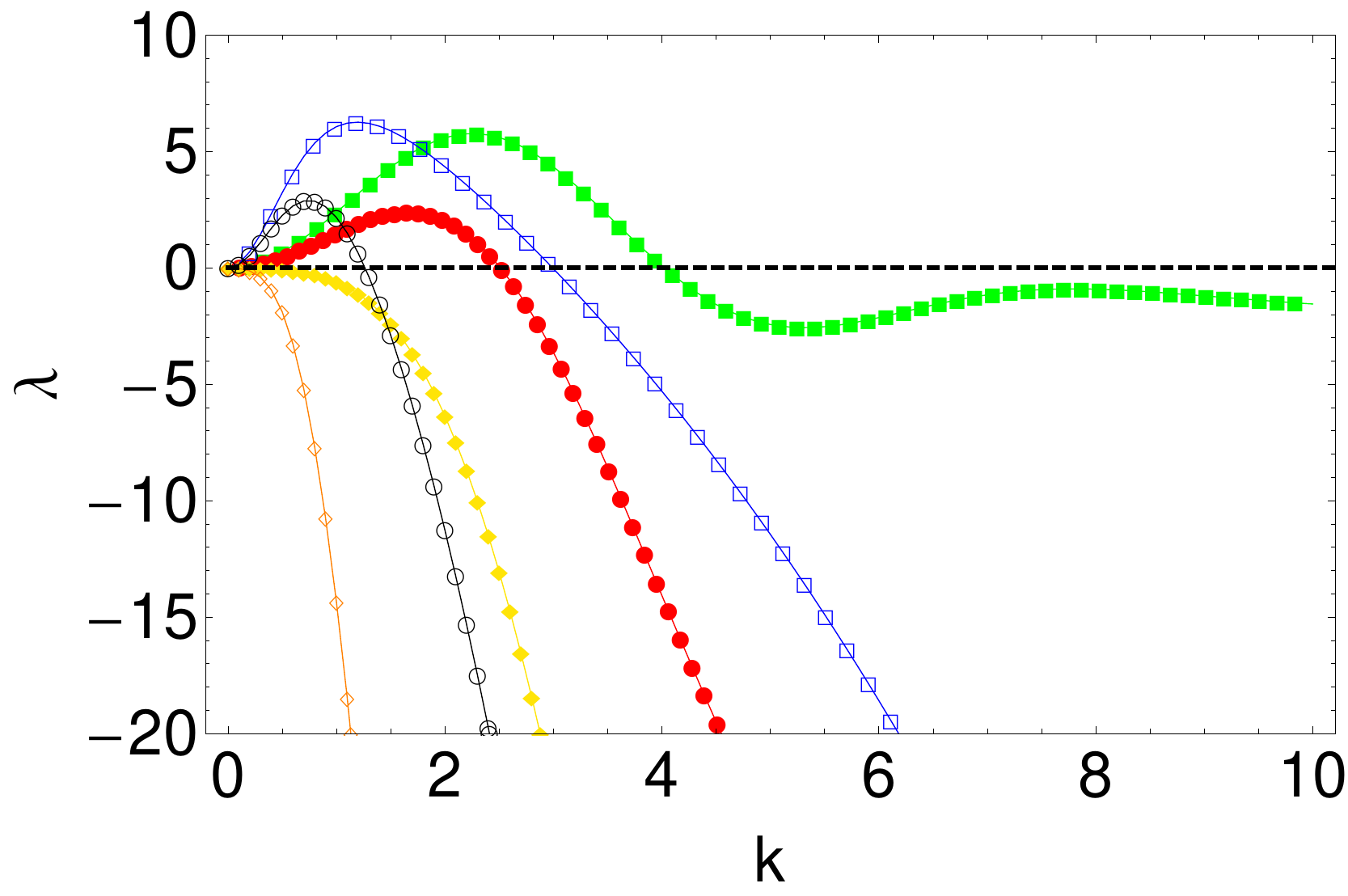}
\caption{\label{fig:stability_attractive}Growth rate (Eq.~(\ref{disp_nonoise}))
for attractive GEM-3 and GEM-1 potentials (full and empty symbols respectively).
For GEM-3: $\tilde{D} = 0.015$ (green squares), $0.9$ (red disks) and $3$ (yellow diamonds).
For GEM-1: $\tilde{D} = 0.6$ (blue squares), $4.5$ (black circles) and $21$ (orange diamonds). }
\end{figure}

A main difference with the repulsive case is that the cluster
patterns appear for any value of $\alpha$. This can be easily
explained from the linear stability analysis of the homogeneous
density. The growth rate of perturbations remains the same as
in Eq.~(\ref{disp_nonoise}) but now, since $\epsilon<0$ the
Fourier transform of the potential will have negative values
independently of $\alpha$. Figure
\ref{fig:stability_attractive} shows the growth rate
$\lambda(k)$ for some parameter values. As before, for large
values of $\tD$, $\lambda (k) <0$ for all $k$, so that no
instability to cluster formation will occur on the homogeneous
density.

\section{Summary and conclusions}
\label{sec:summary}

We have analyzed in detail the properties of a system of
interacting Brownian particles in the presence of a soft-core
repulsive two-body potential. The relevant result is that in a
range of parameters, despite the repulsion, the particles
aggregate in clusters that periodically arrange in space. We
have studied the system at two descriptive levels: the
microscopic particle dynamics, and the DK equation for the
coarse-grained density of particles. By considering the
deterministic version of the DK equation we have obtained the
condition for pattern formation, which is that the Fourier
transform of the potential must have negative values, and when
this is the case, the diffusion coefficient has to be small
enough. When diffusion is small the single clusters have a
Gaussian shape maintained by the interplay between repulsion
between close particles and diffusion, which tend to increase
cluster width, and repulsion from particles in neighboring
clusters, which tends to narrow the clusters. In addition to
clearly identifying the mechanisms involved, our approach based
on the deterministic DK equation has allowed the derivation of
analytical expressions for cluster width and height in 1d and
in 2d which are accurate for small diffusion. The bifurcation
behavior of the steady density has also been analyzed close to
the onset of instability of the homogeneous state, obtaining
approximations for the periodic density patterns formed in 1d
and 2d. Finally, the situation in which the particles interact
attractively has been briefly considered, obtaining also
situations of cluster formation. Similarities and differences
with the repulsive case have been commented.

The consideration of low dimensions (one and two) and the
restriction to a deterministic approach in which pattern
formation techniques become powerful has allowed us to gain
insight into this counterintuitive clustering instability in
which particle repulsion leads to clustering. We close by
noting the strong formal analogies, including the condition for
linear instability, with the situation of cluster formation in
models of population dynamics, with non-conserved number of
particles, in which the repulsive interaction is replaced by a
negative influence of the individuals onto the growth of
others, i.e. competitive interaction
\cite{Hernandez2004,MartinezGarcia2013,Hernandez2014}. The
phenomenon studied here i.e. the formation of crystals of
clusters induced by an instability of the density in the
presence of repulsive or competitive feedbacks is thus very
general and could be found in many other kinds of systems.


 \acknowledgments
We thank Dr. J.J. Cerd\`a for pointing to our attention relevant references. We acknowledge financial support from Ministerio de Econom\'ia y Competitividad and Fondo Europeo de Desarrollo Regional under projects FIS2015-63628-C2-1-R (MINECO/FEDER), FIS2015-63628-C2-2-R (MINECO/FEDER) and CTM2015-66407-P (MINECO/FEDER). C. L. acknowledges the COST Action MP1305, supported by COST (European Cooperation in Science and Technology).



\section*{Appendix: Weakly nonlinear analysis close to instability points}
\label{appendixA}

 We want to obtain an equation for the
deviation of the solution of the Dean equation with respect to
the homogenous solution $\rho_0$ close to its instability
threshold. Using dimensionless variables so that $\rho_0=1$ the
equation for $\delta(\bx,t)=\rho(\bx,t)-1$ is
\BE
\dot \delta(\bx,t) = \tD \nabla^2 \delta(\bx,t) + \bG
\delta(\bx,t) + \nabla \cdot\left( \delta(\bx,t) \bH
\delta(\bx,t) \right) \ .
\EE
We have defined the operators $\bH$ and $\bG$ as
\BE
\bH f(\bx) =\int \nabla \tv(\bx-\bx') f(\bx') d \bx'
\EE
and
\BE
\bG f(\bx) =\int \nabla^2 \tv(\bx-\bx') f(\bx') d \bx'\ .
\EE
We note that
\BE
\bH e^{i\bk\cdot\bx} = i\bk \hv(-\bk) e^{i\bk\cdot\bx} = i\bk
\hv(k) e^{i\bk\cdot\bx}\ ,
\EE
where the last equality arises because $\hv(\bk)$ depends only on
the modulus of $\bk$: $\hv(\bk)=\hv(k)$, and
\BE
\bG e^{i\bk\cdot\bx} = -k^2 \hv(k) e^{i\bk\cdot\bx} \ .
\EE

The homogeneous solution $\delta=0$ becomes unstable for
$\alpha>2$, for which $\hv(\bk)$ has negative Fourier
components, and $\tD<-\hv(k_c)$. We use the notation
$\hv_1=\hv(k_c)$ and $\hv_2=\hv(2k_c)$ and introduce the
expansions
\BA
\tD &=& \tD_c+a_1\eta+a_2\eta^2+a_3\eta^3+\ldots
\label{Dexpan}\\
\delta(\bx,t) &=&
0+\eta\psi_1+\eta^2\psi_2+\eta^3\psi_3 +\ldots
\label{deltaexpan}
\EA
where $\tD_c=-\hv_1=|\hv_1|$, $\psi_i=\psi_i(\bx,T,...)$,
$T=\eta^2 t +\ldots$.

We obtain
\BA
{\cal O} (\eta):   &\bL_c\psi_1 &=0
\label{O1}\\
{\cal O} (\eta^2): &\bL_c\psi_2 &=
-a_1\nabla^2\psi_1-\nabla\cdot\left(\psi_1 \bH \psi_1\right)
\label{O2}\\
{\cal O} (\eta^3): &\bL_c\psi_3 &= \partial_T \psi_1
-a_1\nabla^2\psi_2 -a_2\nabla^2\psi_1 \nonumber \\
&& -\nabla\cdot\left(\psi_1 \bH \psi_2\right) -
\nabla\cdot\left(\psi_2 \bH \psi_1\right) \ , \label{O3}
\EA
where the critical operator is $\bL_c=\tD_c\nabla^2+\bG$.

The general solution of Eq.~(\ref{O1}) is
\BE
\psi_1=\sum_{\bk} A_\bk(T) e^{i\bk\cdot\bx} \ ,
\label{O1sol}
\EE
where the sum is over wavevectors with the critical modulus
$|\bk|=k_c$.

We start with the one-dimensional case, for which Eq.~(\ref{O1sol}) reduces to
\BE
\psi_1=A(T)e^{ik_c x}+\textrm{cc} \ .
\EE
$\textrm{cc}$ means complex conjugate. Substituting in Eq.~(\ref{O2}):
\BE
\bL_c \psi_2 = a_1 k_c^2 A(T) e^{ik_c x} + 2 k_c^2 A(T)^2 \hv_1
e^{2ik_cx} + \textrm{cc} \ .
\EE
Fredholm theorem requires $a_1=0$ to avoid resonances, and
then, neglecting the solution of the homogeneous solution,
$\psi_2$ is given by
\BE
\psi_2= \frac{A(T)^2 e^{2ik_c
x}}{2\left(1-\frac{\hv_2}{\hv_1}\right)} + \textrm{cc} \equiv
B(T) e^{2ik_c x} + \textrm{cc} \ .
\EE
Going to the next order, Eq.~(\ref{O3}):
\BA
&&\bL_c\psi_3 = e^{ik_c x}\left(
\partial_T A(T) +  a_2 k_c^2 A\right. \nonumber \\
 &+& \left.
 k_c^2 B  A^*\left[ 2 \hv_2 - \hv_1 \right] \right) + \textrm{cc} +
{\cal O}(e^{3ik_c x}) \ .
\EA
Again, elimination of resonances requires
\BA
\partial_T A &=& -a_2 k_c^2 A -k_c^2 A^* B \left(2\hv_2-\hv_1\right) \nonumber \\
&=& -a_2 k_c^2 A - k_c^2 \frac{2\hv_2-\hv_1}{2\left(
1-\frac{\hv_2}{\hv_1}\right)} |A|^2 A \ ,
\EA
which is the amplitude equation describing the dynamics at
$\tD\approx \tD_c$. The steady solution is
\BE
|A_{st}|^2=\frac{-2 a_2
\left(1-\hv_2/\hv_1\right)}{2\hv_2-\hv_1} \ .
\EE
Using the expansion (\ref{Dexpan}):
\BE
\eta = \sqrt{\frac{\tD_c-\tD}{-a_2}} + \ldots
\EE
the expansion (\ref{deltaexpan}) for the steady state becomes
\BA
\delta(x)=\left(
\frac{(\tD_c-\tD)2(1-\hv_2/\hv_1)}{2\hv_2-\hv_1}\right) e^{i k_c x +i\phi} + \textrm{cc} \nonumber \\
+ \frac{\tD_c-\tD}{-a_2} \psi_2(x) +
\ldots \label{finaldelta1dApp}
\EA
$\phi$ is the (arbitrary) phase of $A_{st}$, fixing the
position of the pattern, and that in the following we take
$\phi=0$. Finally, using that $D_c=-\hv_1=|\hv_1|$ we find
expression (\ref{amplitude1d}) in the main text. Given the
signs in Eq. (\ref{finaldelta1dApp}), we have a supercritical
bifurcation from the homogeneous state to a periodic array of
clusters when $\tD$ is reduced below $\tD_c$.

The procedure can be repeated in two dimensions. For simplicity
we focus directly on the steady solutions,  so that the term
$\partial_T\psi_1$ is absent from Eq.~(\ref{O3}). The solution
of Eq.~(\ref{O1}) with hexagonal symmetry is
\BE
\psi_1=\sum_{r=1,2,3} A_r e^{i\bk_r \cdot\bx} + \textrm{cc} \ ,
\label{hexagonal}
\EE
where $\bk_r$, $r=1,2,3$ are three wavevectors of modulus $k_c$
and oriented 120 degrees apart. Together with the other three
wavevectors contained in the complex conjugate (cc) terms, they
complete the hexagonal platform that generates the hexagonal
pattern. We note that $\bk_1+\bk_2=-\bk_3$,
$\bk_2+\bk_3=-\bk_1$, and $\bk_3+\bk_1=-\bk_2$. Other vectors
that appear in the nonlinear expansion are
$\bk_{12}=\bk_1-\bk_2$, $\bk_{23}=\bk_2-\bk_3$ and
$\bk_{31}=\bk_3-\bk_1$, all of modulus $k_{12}=\sqrt{3} k_c$.
We define $\hv_{12} \equiv \hv(k_{12})$.

Introducing (\ref{hexagonal}) into the second order
Eq.~(\ref{O2}) we obtain as the conditions for eliminating the
resonant terms:
\BE
a_1 A_1 + \hv_1 A_2^*A_3^*=0 \ ,
\label{FredholmO2}
\EE
and the other two complex equations resulting from cyclic
permutation of the subindices of $A_r$: $1\rightarrow 2
\rightarrow 3 \rightarrow 1$. Choosing appropriately the origin
of coordinates it is enough to solve (\ref{FredholmO2}) for
real and equal amplitudes: $A_r=A$, $r=1,2,3$, so that
\BE
A=\frac{a_1}{|\hv_1|} \ .
\EE
The second order equation can now be solved, giving a $\psi_2$
in terms of $A$ and $a_1$ and with spatial structure containing
wavevectors $2\bk_r$ and $\bk_{rs}$, $r,s=1,2,3$. Eliminating
again resonant terms from Eq.~(\ref{O3}) and defining
$\delta_0=\eta A$ we find the two formulas given in the
main text, namely Eq. (\ref{Kamplitude2d}) and
(\ref{Kamplitude2d_delta}). There is a subcritical bifurcation
from homogeneous to hexagonal density when reducing $\tD$.

\bibliographystyle{siam}
\bibliography{bugsrefs}

\end{document}